\documentclass[conference]{IEEEtran}
\IEEEoverridecommandlockouts
\usepackage{cite}
\usepackage{amsthm}
\usepackage{amsmath,amssymb,amsfonts}
\usepackage{algorithmic}
\usepackage{graphicx}
\usepackage{textcomp}
\usepackage{xcolor}
\usepackage{booktabs}
\usepackage{tabularx}
\usepackage{graphicx}
\usepackage{comment}
\usepackage[ruled,vlined]{algorithm2e}

\newtheorem{proposition}{Proposition}[section]
\newtheorem{theorem}{Theorem}[section]

\def\BibTeX{{\rm B\kern-.05em{\sc i\kern-.025em b}\kern-.08em
    T\kern-.1667em\lower.7ex\hbox{E}\kern-.125emX}}

\begin{document}

\title{UAV-enabled Computing Power Networks: Design and Performance Analysis under Energy Constraints \\ \thanks{A preliminary version of this work has been published in \textit{IEEE GLOBECOM}~\cite{deng2025uav_c}.} 
\thanks{The research described in this paper was conducted in the JC STEM Lab of Smart City funded by The Hong Kong Jockey Club Charities Trust (Contract No. 2023-0108). This work was supported in part by the Hong Kong SAR Government under the Global STEM Professorship and Research Talent Hub. The work of Yiqin Deng was supported in part by the National Natural Science Foundation of China (Grant No. 62301300) and by the Shandong Provincial Natural Science Foundation (Grant No. ZR2023QF053). The work of Senkang Hu was supported in part by the Hong Kong Innovation and Technology Commission under InnoHK Project CIMDA. The work of Haixia Zhang was supported in part by the Joint Funds of the NSFC (Grant No. U22A2003). The work of Yuguang Fang was also supported in part by a grant from the Research Grants Council of the Hong Kong Special Administrative Region, China (Project No. CityU 11216324). (Corresponding author: Yuguang Fang) }
\thanks{Yiqin Deng is with School of Data Science, Lingnan University, Tuen Mun, Hong Kong, China (email: yiqindeng@ln.edu.hk).}
\thanks{Zhengru Fang, Senkang Hu, Yanan Ma, and Yuguang Fang are with Hong Kong JC STEM Lab of Smart City and Department of Computer Science, City University of Hong Kong, Kowloon, Hong Kong, China (email: \{zhefang4-c, senkang.forest, yananma8-c\}@my.cityu.edu.hk, my.fang@cityu.edu.hk).}
\thanks{Xiaoyu Guo is with Department of Biomedical Engineering, City University of Hong Kong, Kowloon, Hong Kong, China (e-mail:
xiaoyguo@cityu.edu.hk).}
\thanks{Haixia Zhang is with the Institute of Intelligent Communication Technology, and also with the Shandong Key Laboratory of Intelligent Communication and Sensing-Computing Integration, Shandong University, Jinan 250061, Shandong, China (email: haixia.zhang@sdu.edu.cn).}
}

\author{\IEEEauthorblockN{Yiqin Deng, Zhengru Fang, Senkang Hu, Yanan Ma, Xiaoyu Guo, Haixia Zhang, and Yuguang Fang,~\IEEEmembership{Fellow,~IEEE}
}
}

\maketitle
\thispagestyle{plain}

\begin{abstract}
This paper presents an innovative framework that boosts computing power by utilizing~\textit{ubiquitous computing power distribution} and enabling higher~\textit{computing node accessibility} via adaptive UAV positioning, establishing a~\textit{UAV-enabled Computing Power Network (UAV-CPN)}.
In a UAV-CPN, a UAV functions as a dynamic relay, outsourcing computing tasks from the~\textit{request zone} to an expanded~\textit{service zone} with diverse computing nodes, including vehicle onboard units, edge servers, and dedicated powerful nodes. This approach has the potential to alleviate communication bottlenecks and overcome the ``island effect'' observed in multi-access edge computing.
A significant challenge is to quantify computing power performance under complex dynamics of communication and computing. To address this challenge, we introduce~\textit{task completion probability} to capture the capability of UAV-CPNs for task computing. We further enhance UAV-CPN performance under a hybrid energy architecture by jointly optimizing UAV altitude and transmit power, where fuel cells and batteries collectively power both UAV propulsion and communication systems. Extensive evaluations show significant performance gains, highlighting the importance of balancing communication and computing capabilities, especially under dual-energy constraints. These findings underscore the potential of UAV-CPNs to significantly boost computing power. 
\end{abstract}


\section{Introduction}
\label{sec:intro}
The proliferation of latency-sensitive and computation-intensive applications, such as augmented reality and autonomous driving~\cite{Fang2025, Hu2025}, has intensified the demand for ubiquitous connected networks that seamlessly integrate communication and distributed computing capabilities~\cite{Xiao2024Space}. While multi-access edge computing (MEC) architectures have expanded computing capability at the network edge~\cite{deng2022actions,ma2025raise}, existing computing resources remain geographically fragmented, leading to resource under-utilization and creating isolated computing power islands, known as ``island effect''~\cite{Sun2024}. The
computing power network (CPN) initiative aims to interconnect these islands via computation-aware networking, but faces challenges in accessibility, cost-efficiency, and resilience,
especially during disasters or traffic peaks~\cite{tang2021computing,Sun2024}. Specifically, one major challenge is the communication bottleneck in providing  ground users (GUs) with viable access to computing power  on a wide range of mobile devices and in accessing isolated computing power islands (e.g., edge servers, computing clusters, and user-provided computing nodes). Another challenge is the cost associated with the ubiquitous deployment of computing nodes (CNs), and it is not cost-effective to deploy fixed edge servers to handle infrequently occurring local computing demands~\cite{Sun2024, ma2025uav}. In addition, during events such as disasters or rush hours~\cite{dai2024uav,hao2024joint}, where infrastructure may fail or computing resources become insufficient, these challenges are further exacerbated. Inaccessibility to computing resources can have devastating consequences~\cite{Basu2019a}.

Unmanned aerial vehicles (UAVs) offer a promising solution due to their rapid deployment, flexible 3D mobility, line-of-sight (LoS) connectivity, and on-demand provisioning of communication and computing services~\cite{Zhan2018Energy, Li2023Channel, Xiao2024Space, Li2025Large}.  UAV-assisted MEC has been widely studied~\cite{Dong2024, liu2025multi,wu2025joint,xiao2025star,Telikani2025unmanned, tao2024multi,ma2025uav}, typically under the assumption of \emph{static CN accessibility}, where the number and locations of available CNs are fixed and independent of a UAV's position. This rigid assumption fundamentally limits scalability and fails to exploit the vast pool of spatially distributed computing resources (e.g., edge servers, connected autonomous vehicles, and user devices) that may become accessible under favorable UAV deployment and latency conditions~\cite{deng2024uav}. As a result, existing models cannot fully capture the opportunistic nature of resource integration in large-scale or infrastructure-scarce environments.

In contrast, we propose a novel paradigm: \emph{UAV-enabled Computing Power Networks (UAV-CPNs)}, where UAVs serve as dynamic aerial relays that adaptively connect spatially distributed and heterogeneous computing resources on demand into a unified service network architecture. Particularly, the set of accessible CNs is not predetermined; instead, it is dynamically shaped by a UAV's 3D positioning and the end-to-end (E2E) latency requirements of tasks. This dynamic accessibility to computing resources enables more efficient utilization of ubiquitous but spatially distributed dynamic computing resources, especially in large-scale or infrastructure-scarce scenarios.

However, this flexibility introduces new modeling and optimization challenges. For instance, UAV altitude affects multiple performance dimensions. On the one hand, higher altitudes increase the existence probability of LoS links between GUs and the UAV, improving uplink reliability for \textit{GU-to-UAV task offloading} and extending downlink reachability to remote CNs. On the other hand, increased altitude also leads to higher path loss and transmission delay due to longer propagation distances, degrading signal quality and increasing communication latency. 
Consequently, the optimal altitude that maximizes uplink rates may be suboptimal for \textit{UAV-to-CN forwarding} and hence for overall task completion, as downlink performance depends on UAV-CN channel quality and CN accessibility, both of which are particularly sensitive to UAV positioning. These interdependent trade-offs create a complex optimization landscape, particularly in the vertical dimension, which remains underexplored in the current literature. In this paper, we therefore take the altitude as the primary geometric degree of freedom, while modeling the horizontal spatial variability of GUs and CNs statistically via homogeneous point processes. This leads to a horizontally homogeneous large-scale network model in which the absolute horizontal coordinates of the UAV are abstracted into spatial distributions, and the altitude emerges as the key deterministic parameter that shapes the GU–UAV and UAV–CN distance distributions, LoS/NLoS probabilities, and dual-energy consumption. Our focus on the vertical dimension provides a first-step towards the theoretical understanding of UAV-CPNs and is complementary to existing studies that fix the UAV altitude and optimize only the horizontal placement or trajectory in finite-user scenarios.

In addition to this complexity, practical UAV-CPNs must support extended operational durations under high service demands. Hybrid fuel cell and battery-powered systems, such as hydrogen fuel cell powered UAVs~\cite{Guo2025integrated}, are increasingly favored over conventional battery-only UAVs for their higher energy density and longer endurance~\cite{zhao2025dynamic}. In such hybrid architectures, the fuel cell and battery are typically integrated in serial, parallel, or decoupled configurations, each with distinct energy management strategies. The fuel cell, benefiting from its high energy density, serves as the primary power source for sustained propulsion and may also recharge the battery in certain designs. In contrast, the battery acts as a secondary energy buffer, delivering high-power bursts to support communication and control subsystems.  
This dual-power architecture enables UAV-CPNs to simultaneously support prolonged flight and intensive task computing. However, it also introduces \emph{dual-energy constraints}: the task may fail if either the fuel cell energy or the battery capacity is depleted.
 In contrast with conventional battery-powered UAV networks~\cite{xu20243d, gong2024energy, zhu2024multi, Nguyen2025integrated}, where propulsion dominates energy consumption and communication energy is often negligible, hybrid systems require fine-grained coordination between the two energy sources. In this context, communication energy consumption becomes a critical factor and can no longer be ignored. As a result, performance bottlenecks may arise in communication, computing, or energy domain (propulsion or communication power), potentially leading to task failure. This necessitates a holistic optimization framework in which both UAV deployment and resource allocation are jointly designed to maximize system performance, considering not only computing task scheduling but also the efficient management of dual-energy supplies.

In summary, this is the first work to model and optimize UAV-CPNs under dynamic CN accessibility and dual-energy constraints. The main contributions are summarized as follows:
\begin{itemize}
\item\textbf{A novel UAV-CPN framework:} We propose a spatially dynamic model where the set of accessible CNs is determined by a UAV's positioning and task latency requirements. This model enables opportunistic integration of geographically distributed computing resources, breaking the ``island effect'' of conventional static-access computing architectures. Focusing on a foundational scenario involving a single UAV and its vertical positioning, we define~\textit{task completion probability} as the critical performance metric and develop an analytical procedure to obtain the performance metric. 
    
\item\textbf{Joint optimization under dual-energy budgets:} For practical hybrid fuel cell and battery-powered UAV scenarios, we introduce novel propulsion and communication energy models for UAV-CPNs. Under these dual-energy models, we formulate a task completion probability maximization problem subject to both fuel (for propulsion) and battery (for communication) energy constraints. Moreover, we design a computationally efficient algorithm to jointly optimize UAV transmit power and altitude.
    
    \item \textbf{Performance evaluation and insights:} Through extensive numerical analysis, we verify the accuracy of our analytical model and uncover trade-offs between communication parameters (e.g., UAV altitude and transmit power) and computing parameters (e.g., CN density and coverage), highlighting potential bottlenecks in resource allocation. Under energy constraints, we quantify the performance gains from joint power-altitude optimization compared to single-parameter optimization and static strategies. These findings provide concrete guidelines for network deployment.
\end{itemize}

The remainder of this paper is organized as follows. Section II reviews related works and Section III introduces the proposed architecture. Section IV develops an analytical framework for key performance metrics. Building on this foundation, Section V analyzes system performance under energy constraints and presents solutions for joint optimization. Section VI provides comprehensive numerical evaluations to validate theoretical findings and quantify performance gains through parameter optimization. Finally, Section VII concludes the paper.

\section{Related Works}
In this section, we only review research works closely related to this paper in two aspects: UAV-assisted MEC frameworks and energy-constrained UAV optimization.

\subsection{UAV-assisted Computing Framework}
Numerous studies have focused on user association, computation offloading, and resource allocation in various UAV-assisted MEC systems with UAVs serving as relay nodes~\cite{qi2024minimizing,lu2024resource,pan2024resource}, aerial edge servers~\cite{lin2024a,liu2025multi,wu2025joint,lyu2025empowering}, and/or both~\cite{wang2024joint,xiao2025star,Nabi2025joint}. These works have addressed design challenges, including energy consumption minimization, latency reduction, and task throughput enhancement. 
Recent research has also explored UAV-assisted CPNs, which is essentially a form of traditional UAV-assisted MEC systems. These CPNs are characterized by greater heterogeneity and dynamics in their computing resources, encompassing CPUs, GPUs, and TPUs, respectively~\cite{tao2024multi,ma2025uav}. However, most of these studies assume that the set of available CNs remains fixed and predetermined, regardless of a UAV’s deployment location or the spatial distribution of computing resources. This static CN accessibility model significantly limits the system’s ability to leverage computing resources that are spatially distributed outside the immediate task request area. As a result, it leads to a mismatch between computing demand and supply, especially in scenarios where the local computing infrastructure is insufficient or unavailable~\cite{Deng2024}. This limitation is particularly critical in post-disaster situations, where UAVs must dynamically access computing resources outside the damaged geographic area to support mission-critical tasks.

In contrast to these works, our model does not assume a fixed set of CNs. Instead, the effective set of computing nodes accessible to UAVs is dynamically determined by their deployment position and the task’s E2E latency requirement. This approach enables UAVs to opportunistically access distributed computing resources across a large spatial area, thereby enhancing the flexibility and efficiency of UAV-assisted computing systems.

\subsection{Energy-Constrained UAV Optimization}
Many research works also investigate methods to improve system performance, including perception, communication, and computing in UAV-assisted wireless networks under energy constraints through 3D trajectory design and resource allocation~\cite{Zeng2017Energy, Xu2018UAV, Zhan2018Energy, Zeng2019Energy, xu20243d,gong2024energy,zhu2024multi,Nguyen2025integrated}. In these studies, typical electric  (battery-only) UAVs are adopted to support system operations, where the energy consumed in propelling UAVs significantly dominates that consumed in communication or/and computation, making propulsion the primary energy consumer. Indeed, many existing studies neglect transmission power altogether~\cite{zhang2020energy}. Although in \cite{lin2024a}, Lin~\textit{et al.} consider using a solar-powered UAV as an edge server to perform data collection and processing, where the amount of the harvested solar energy increases with flying altitude, but they only address the tradeoff between energy harvesting and communication performance. Similar to other studies, communication-related energy consumption has not been taken into consideration.
With the advent of hybrid fuel cell and battery-powered UAVs, such as hydrogen fuel cell powered UAVs~\cite{Guo2025integrated}, a novel paradigm has emerged where both fuel cells and batteries are utilized to power propulsion, communication, and control modules. Depending on the energy management topology, whether the fuel cell and battery are configured in series or in parallel, the efficiency of energy utilization can vary~\cite{zhao2025dynamic}. Unlike traditional battery-powered UAVs, in this new paradigm, transmission energy consumption becomes crucial for system sustainability and must be considered alongside UAV deployment and resource allocation. Failure in either energy source, fuel or battery, can lead to task failure.
Building on this, Zhang et al.~\cite{zhang2020energy} defined the energy efficiency in a hybrid fuel-powered UAV-relay transmission system as the ratio of transmitted data to total energy consumption, optimizing UAV trajectories and node transmission power to maximize the average energy efficiency during each time slot. However, no prior work has investigated computing performance within hybrid fuel cell and battery-powered UAV systems, where computing capability and communication conditions are closely intertwined with UAV deployment and resource allocation.

Compared to existing studies, this paper is the first to consider the use of UAVs to enable broader accessibility of potentially remote ubiquitous computing nodes both within and beyond the service request area (i.e., the \textit{request zone}), thereby optimizing task computing efficiency. Considering the duration requirements of UAV operations, we adopt hybrid fuel cell and battery-powered UAVs, where both propulsion and communication energy consumption are jointly considered. This approach consequently introduces a unique challenge of balancing the communication-computing-energy tradeoff. Failure to address this balance could create systematic bottlenecks, leading to inefficiencies or failures in task computing.

\section{Modeling UAV-CPNs}
\label{sec:system_model}
In this section, we present the foundational framework for UAV-CPNs by modeling the core components: the network spatial model, air-ground channel model, and computing model. These models collectively provide the analytical basis for subsequent performance analysis.

To simplify the analysis for analytical tractability, we adopt a single-UAV with a single-CN system, which is to establish a theoretical performance floor for the system, providing a conservative performance benchmark that deliberately excludes performance gains from multi-CN parallel processing, a common optimization strategy in existing literature~\cite{Deng2024}. The proposed UAV-CPN framework itself is not restricted to this single-CN setting and, in principle, supports multi-CN parallel and cooperative processing; a more comprehensive treatment of joint CN selection and task partitioning will be left for future work.

\begin{figure}[t]
    \centering
    \includegraphics[width=0.825\linewidth]{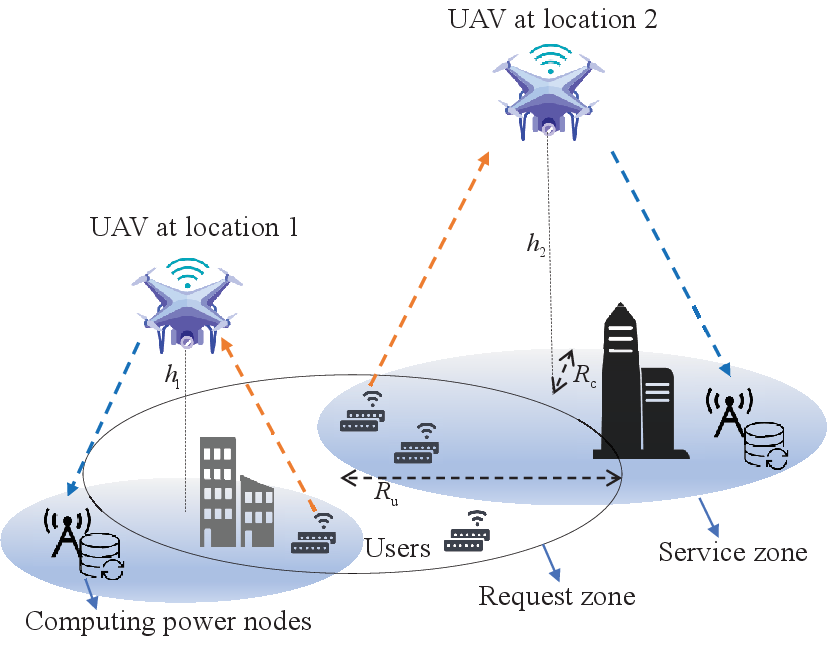}
    \caption{Illustration of a UAV-enabled computing power network, where GUs offload tasks generated within the service area to distributed computing power nodes for processing. The computing accessibility is enhanced by strategically adjusting key network parameters, e.g., the position of the aerial UAV relay. }
    \label{fig:system}
\end{figure}

\subsection{Network Spatial Model}
To establish a tractable analytical framework for UAV-CPNs, we consider a single UAV-assisted computing scenario as shown in Fig.~1. A circular region of radius $ R_u $, referred to as the \textit{request zone}, contains GUs that generate computation tasks. The locations of GUs are modeled as an independent uniform point process within the \textit{request zone}.

Computing nodes (CNs), on the other hand, are distributed over an unbounded plane, referred to as the \textit{service zone}, and are assumed to follow a homogeneous spatial Poisson Point Process (PPP) $ \Phi_C $ with density $ \lambda_c $ (nodes/m$^2$). The spatial distributions of GUs and CNs are mutually independent, reflecting the independence between task generation and CN availability. Unlike conventional works that assume a fixed number and known locations of CNs, our proposed model reflects the spatial randomness and ubiquity of distributed computing resources. Importantly, the set of CNs accessible to the UAV is not fixed in advance, but is dynamically determined based on the UAV’s deployment position and the task’s E2E latency requirements.

We assume that a hybrid fuel cell and battery-powered rotary-wing UAV is deployed at  altitude~$ h $ above the center of the \textit{request zone}. It acts as an aerial relay to support two-way computation offloading between GUs and remote CNs. The effective \textit{service zone} accessible by a UAV is determined by both the UAV’s communication range and the E2E latency requirements. Specifically, only those
CNs that can satisfy the E2E latency constraints are considered valid candidates for task forwarding.

Let the UAV-to-CN distance be denoted as $ d_{u,c} $, and the GU-to-UAV distance as $ d_{g,u} $. The total E2E latency for a task from a GU to a selected CN via the UAV is given by:
\begin{equation}
T_{\text{total}} = T_{\text{offload}} + T_{\text{forward}} + T_{\text{compute}} + T_{\text{return}},
\end{equation}
where $ T_{\text{offload}} $ is the GU-to-UAV transmission time, $T_{\text{forward}}$ is the UAV-to-CN transmission time, $T_{\text{compute}}$ is the CN computing time,
 and $T_{\text{return}}$ is the result return time from CN to GU. To ensure quality-of-service (QoS) compliance, we define a maximum allowable latency budget $ T_{\max} $ such that
\begin{equation}\label{eq:constraint}
T_{\text{total}} \leq T_{\max}.
\end{equation}
This constraint limits the maximum UAV-to-CN distance (i.e., the maximum \textit{service zone} radius) $ d_{u,c}^{\max} $, and consequently defines the effective size of the \textit{service zone} accessible by the UAV. Specifically, $ d_{u,c}^{\max} $ is determined by the worst-case combination of $ d_{g,u} $ and $ d_{u,c} $ under the  constraint in Eq.~\eqref{eq:constraint}.

Under the PPP assumption, the expected number of CNs within a subregion $ \mathcal{A} $ of the \textit{service zone} is~\cite{Møller_Schoenberg_2010}:
\begin{equation}
\mathbb{E}[N_{\text{CN}}(\mathcal{A})] = \lambda_c \cdot |\mathcal{A}|,
\end{equation}
where $ |\mathcal{A}| $ denotes the area of $ \mathcal{A} $. Similarly, the expected number of GUs in a subregion of the \textit{request zone} is determined by the GU density and the subregion area.
At the framework level, all CNs located within the \textit{service zone} and satisfying the latency constraint in (\eqref{eq:constraint}) are treated as potential candidates for serving offloaded tasks, so that a GU is not restricted to receiving service from a pre-specified single node. In the subsequent analysis, we will specialize this general spatial model to a simplified single-CN benchmark scenario in order to develop a tractable performance analysis framework and gain fundamental insights into the system behavior.

\subsection{Air-ground Channel Model}
Following~\cite{al2014optimal}, we adopt a probabilistic air-ground channel model to characterize the mixture of line-of-sight (LoS) and non-line-of-sight (NLoS) channels for both the \textit{GU-to-UAV task offloading} link and the \textit{UAV-to-CN task forwarding} link. This model is widely used in the current literature on UAV communications to capture the aggregate impact of multipath propagation in diverse complex urban environments, where signal reflections, diffractions, and scattering significantly affect transmission quality, while still providing a tractable basis for performance analysis.

In the \textit{GU-to-UAV task offloading} phase, the probability that the link is LoS is given by~\cite{al2014optimal}:
\begin{equation}
     P_{\text{LoS,up}} = \frac{1}{1 + C \exp\left(-B\left(\frac{180}{\pi}\arctan\left(\frac{h}{r_u}\right) - C\right)\right)},
\end{equation}
where $ B $ and $ C $ are environment-dependent parameters that reflect the urban or rural landscape's impact on LoS existence probability. The GU-to-UAV channel model is then characterized as:
\begin{equation}\label{eq:up_power}
     P_{r,\text{up}} = \begin{cases} 
     P_u \cdot \left(r_u^2 + h^2\right)^{-\alpha_u/2} & \text{with}\, P_{\text{LoS,up}}, \\
     \eta P_u \cdot \left(r_u^2 + h^2\right)^{-\alpha_u/2} & \text{with}\,1-P_{\text{LoS,up}},
     \end{cases}
\end{equation}
where $P_u$ is the GU transmit power, $r_u$ is the horizontal GU--UAV distance, $\alpha_u$ is the uplink path-loss exponent, and $\eta \in (0,1)$ is the NLoS attenuation factor that accounts for additional signal degradation due to multipath effects such as scattering and reflection.

Analogously, for the \textit{UAV-to-CN task forwarding}, the received power at a CN, denoted as $ P_{r, \text{down}} $, can be modeled based on the transmit power of the UAV $ P_d $, the horizontal UAV-CN distance $ r_c $, and the downlink path loss exponent $ \alpha_c $. Thus, the UAV-to-CN channel can be characterized similarly to the GU-to-UAV channel as in Eq.~\eqref{eq:up_power}.

The proposed framework is also compatible with more general wireless fading models. While the current analysis focuses on the above probabilistic LoS/NLoS air–ground channel model for analytical clarity, it can be extended to incorporate small-scale fading effects such as Rayleigh, Rician, or Nakagami fading by introducing random channel gain terms in Eq.5. For a given fading model, these gains follow known distributions, and the task completion probability expressions developed in Section~IV retain the same functional structure, with the deterministic received-power terms replaced by their averages over their fading distribution. This leads to one-dimensional (or low-dimensional) integrals that can be efficiently evaluated numerically within our semi-analytical performance analytic framework, instead of closed-form solutions. Therefore, extending the channel model to include more general fading remains computationally feasible and does not change the order of complexity of the proposed derivation and optimization algorithm, while allowing the framework to capture both large-scale path loss and small-scale signal fading due to multipath propagation and thereby enhancing its applicability to diverse wireless environments. 

\subsection{Computing Model}
We consider heterogeneous CNs cooperatively operated by multiple service providers or CN node owners, where some factors such as hardware heterogeneity, queuing delays, and I/O interference among virtual machines collectively influence computational throughput~\cite{Huang2018}. To generically model the dynamic computing capabilities of CNs, reflecting their uncertainty as well, we characterize the computation time \( t_c \) through its cumulative distribution function (CDF):
\begin{equation}\label{eq:comp_latency}
F_{t_c}(t; D) = \mathbb{P}(t_c \leq t), 
\end{equation}
where $t$ represents the available time budget for task completion, and \( D \) denotes the task-specific computational workload. This CDF quantifies the probability that a CN completes the task within the given time budget $t$ with workload $D$. 

\section{Task completion probability analysis}
To derive tractable analytical insights from the general UAV-CPN model introduced in Section~\ref{sec:system_model}, in this section, we focus on a simplified baseline scenario where a UAV deployed at altitude \( h \) above the centroid of a circular \textit{request zone} with radius \( R_u \) forwards a typical single GU's computational tasks (with data size \( D \)) to a \emph{single} randomly selected CN within the \textit{service zone}.

This single-CN specialization is adopted solely for analytical tractability and to establish a theoretical performance floor for the system, providing a conservative performance benchmark that deliberately excludes performance gains from multi-CN parallel processing, a common optimization strategy in the existing literature~\cite{Deng2024}. The proposed UAV-CPN framework itself is not restricted to this single-CN setting and, in principle, supports multi-CN parallel and cooperative processing; a more comprehensive treatment of joint CN selection and task partitioning is left for future work.

To investigate the fundamental communication-computing tradeoff, we adopt idealized conditions without consideration of multi-user interference and resource contention, as commonly done in the current literature.  
We begin with modeling the E2E latency for task computing in our UAV-CPN framework. Building on this model, we formally define the task completion probability and derive a semi-analytical expression for a GU at an arbitrary location within the \textit{request zone}. We then obtain the average value of this metric by spatially averaging over all GU positions to characterize system-wide performance, which is equal to the task completion rate or task throughput defined as in MEC systems~\cite{deng2022actions,deng2024uav
}. Although closed-form expressions in elementary functions are generally unavailable, the resulting one- and two-dimensional integrals can be efficiently evaluated numerically within our performance analytic framework. Through this semi-analytical analysis, we reveal the fundamental trade-offs among several critical parameters such as CN density \( \lambda_c \), UAV operational altitude \( h \), and latency budget \( T_{\text{max}} \).
The numerical results based on this framework demonstrate how these parameters jointly determine the operational success of UAV-CPN task computing, yielding actionable design insights for latency-sensitive applications.

\subsection{End-to-end latency}
Given the bandwidth $W$ and noise power $N_0$, the transmission latency from a GU to the UAV ($t_1$) can be calculated by:
\begin{equation}\label{eq:t1}
t_1 = \frac{D}{W \log_2 \left(1 + {P_{r,\text{up}}}/{N_0}\right)}.
\end{equation}
Similarly, the transmission latency from the UAV to a CN ($ t_2 $) can be calculated by:
\begin{equation}\label{eq:t2}
t_2 = \frac{D}{W \log_2 \left(1 + {P_{r,\text{down}}}/{N_0}\right)}.
\end{equation} 
The E2E latency represents the total time from task generation at the GU to the result reception. As assumed in most prior works, we ignore result feedback latency due to the small size of the result, yielding the total E2E latency as
$T_{\mathrm{E2E}}=t_1+t_2+t_c$.

\subsection{Performance metric}
The task completion probability serves as a critical performance metric, reflecting both the communication and computational capabilities of the system, which can be used to compute other performance metrics. For a specific GU, we define the task completion probability as the likelihood of locating a CN within the~\textit{service zone} to complete the computing task at this CN within the E2E latency constraint. Specifically, this means that the condition $ t_1 + t_2 + t_c \leq T_{\text{max}} $ is satisfied. For system-wide analysis, this metric should be spatially averaged over all GU and CN positions governed by their respective distributions. 

If \( t_1 + t_2 \geq T_{\text{max}} \), the accumulated transmission latency across the \textit{GU-to-UAV task offloading} and \textit{UAV-to-CN task forwarding} phases exceeds the latency budget, resulting in a communication bottleneck, termed the \textit{comm-limited} scenario. Instead, if \( t_c \geq T_{\text{max}} - t_1 - t_2 \), the computational latency at the CN exceeds the residual time budget \( T_{\text{res}} \), resulting in a computational bottleneck termed the \textit{comp-limited} scenario.  
Task completion probability necessitates the concurrent fulfillment of both communication and computing constraints. Violation of either constraint, whether \textit{comm-limited} or \textit{comp-limited}, severely degrades system performance. Leveraging stochastic geometry, we derive analytical expressions for the task completion probability, which is the computing power performance metric we will use to derive other performance metrics. 

\subsection{Bottleneck Analysis}
In this subsection, we derive the task completion probability. For a GU at horizontal distance \( r_u \) from the UAV, the latency constraint \( T_{\text{max}} \) fundamentally limits service accessibility despite the theoretical availability of all CNs in UAV-CPNs. Specifically, the \textit{comm-limited} condition imposes a critical spatial restriction by bounding the maximum   radius \( r_c^{\mathrm{max}}(r_u) \) for the effective \textit{service zone}, which is determined by:
\begin{equation}\label{eq:max_coverage}
t_2=t_2\left(r_c^{\mathrm{max}}(r_u)\right) = T_{\text{max}} - t_1(r_u),
\end{equation}
where $t_1=t_1(r_u)$ and $t_2=t_2(r_c)$ are the transmission times for~\textit{GU-to-UAV task offloading} and~\textit{UAV-to-CN task forwarding}, respectively.

Within this communication-constrained~\textit{service zone} (\( r_c \leq r_c^{\mathrm{max}}(r_u) \)), CNs must additionally satisfy the computing latency requirement. For residual time budget \( T_{\text{res}} \triangleq T_{\text{max}} - t_1 - t_2 \), the probability of a CN satisfying this latency constraint is quantified through its CDF as \( F_{t_c}(T_{\text{res}}; D) \), as characterized by Eq.~\eqref{eq:comp_latency}.  

Based on the previous analysis, we establish that task completion fails under either \textit{comm-limited} or \textit{comp-limited} conditions. Specifically, transmission failure occurs when CNs reside outside the communication-constrained~\textit{service zone} (\( r_c \leq r_c^{\mathrm{max}}(r_u) \)), leading to failure in task computing during the transmission phase. For successfully transmitted tasks, the task completion probability equals the likelihood of satisfying the computing latency requirement within the residual time budget $T_{\text{res}}$. These dual constraints induce a \textit{probabilistic thinning}~\cite{Lalley20XX} of the original homogeneous PPP \( \Phi_c \), producing a thinned PPP with spatially varying density. The resulting effective CN density is formally characterized in Proposition~\ref{prop:density}.

\begin{proposition}[Effective CN Density]
\label{prop:density}
For a GU at a horizontal distance \( r_u \) from the UAV, the spatial density of CNs satisfying E2E latency constraint, simply called effective density or conditional spatial density, is given by:
\begin{equation}
\lambda_c^{\mathrm {eff}}(r_u) =\lambda_c  \mathbb{I}_{\{r_c \leq r_c^{\mathrm{max}}(r_u)\}} \cdot F_{t_c}(T_{\text{res}}; D),
\end{equation}
where \( \mathbb{I}_{\{r_c \leq r_c^{\mathrm{max}}(r_u)\}} \) indicates successful communication; \( F_{t_c}(T_{\text{res}}; D) \), as the CDF of computing latency, quantifies the probability of computing latency within $T_{\text{res}}$.
\begin{proof}
A qualified CN must be located within the maximum coverage radius \( r_c^{\mathrm{max}}(r_u) \) for the \textit{service zone} and have a computing latency \( t_c \leq T_{\text{res}} \). The former condition is modeled by an indicator function, referred to as \textit{success in communication}. The latter models the probability that a CN satisfies the computing latency constraint, referred to as \textit{success in computing}. Since the spatial distribution and computing power dynamics are independent, the density of qualified CNs for task completion is the product of these two quantities. Thus, each CN at distance $r_c$ is retained with probability:
\begin{equation}
p_{\text{retain}}(r_u) = 
\underbrace{\mathbb{I}_{\{r_c \leq r_c^{\mathrm{max}}(r_u)\}}}_{\text{Success in communication}} 
\cdot 
\underbrace{ F_{t_c}(T_{\text{res}}; D)}_{\text{Success in computing}}.
\end{equation}
This qualification process represents an \textit{independent thinning} of the original PPP $\Phi_c$. By the Poisson thinning property~\cite{Lalley20XX}, the resulting spatially thinned process retains the PPP property with effective density:
\begin{equation}
\lambda_c^{\mathrm{eff}}(r_u) = \lambda_c \cdot p_{\text{retain}}(r_c).
\end{equation}
\end{proof}
\end{proposition}
Since the thinning process preserves the PPP properties, the qualified CNs form a thinned PPP with effective density \(\lambda_c^{\mathrm{eff}}(r_u)\). The expected number of qualified CNs is derived by integrating this effective density over the spatial domain of interest, as formalized in Proposition~\ref{prop:number}. 

\begin{proposition}[The Qualified Number of CNs]
\label{prop:number}
The expected number of qualified CNs for a GU located at a horizontal distance $ r_u $ from the UAV is given by:
\begin{equation}\label{eq:intensity}
\Lambda(r_u) = 2\pi\lambda_c \int_0^{r_c^{\mathrm{max}}(r_u)}  F_{t_c}(T_{\text{res}}; D)r_c \, dr_c.
\end{equation}
\begin{proof}
The expected number of qualified CNs is derived from the intensity measure of the thinned PPP:
\begin{equation}\label{eq:integral}
\Lambda = \iint_{\mathbb{R}^2} \lambda_c^{\mathrm{eff}}(\|\boldsymbol{x}\|) d\boldsymbol{x}.
\end{equation}
Exploiting circular symmetry about the UAV, we convert to polar coordinates ($d\boldsymbol{x} = r_c dr_c d\theta$):
\begin{equation}
\begin{aligned}
\Lambda(r_u) &= \int_0^{2\pi} \int_0^{r_c^{\mathrm{max}}(r_u)} \lambda_c^{\mathrm{eff}}(r_c) r_c \, dr_c \, d\theta \\
&= 2\pi \lambda_c \int_0^{r_c^{\mathrm{max}}(r_u)}  F_{t_c}(T_{\text{res}}; D) r_c \, dr_c.
\end{aligned}
\end{equation}
\qedhere
\end{proof}
\end{proposition}

\subsection{Task completion probability}
Based on the previous analysis, the task completion probability for a specific GU is summarized in  Theorem~\ref{theorem:success}.
\begin{theorem}\label{theorem:success}
For a GU located at a horizontal distance $ r_u $ from the UAV, the probability that at least one CN satisfying the E2E latency constraint $ t_1 + t_2 + t_c \leq T_{\text{max}} $ can be found is given by:
\begin{multline}\label{eq:success}
P_{\text{success}}(r_u) = 1 - \exp\biggl( -2\pi \lambda_c \\
\times \int_0^{r_c^{\mathrm{max}}(r_u)}  F_{t_c}\left(T_{\text{max}} - t_1(r_u) - t_2(r_c); D\right) r_c \, \mathrm{d}r_c \biggr),
\end{multline}
where $ t_1(r_u) $, $ t_2(r_c) $, $ r_c^{\mathrm{max}}(r_u) $, and $ F_{t_c}\left(T_{\text{max}} - t_1(r_u) - t_2(r_c); D\right)$ are obtained from Eqs.~\eqref{eq:up_power}-\eqref{eq:max_coverage} when the GU is located at a horizontal distance $ r_u $ from the UAV.

\begin{proof}
From Proposition~\ref{prop:number}, the number of qualified CNs, for a GU located at a horizontal distance $ r_u $ from the UAV, follows a Poisson distribution with mean \( \Lambda(r_u) \). The void probability (i.e., no qualified CNs exist) is \( \exp(-\Lambda(r_u)) \)~\cite{Møller_Schoenberg_2010}. Thus, the success probability, which is the probability that there exists at least one CN that can complete the task, is given by: $P_{\text{success}}(r_u) = 1 - \exp(-\Lambda(r_u))$. Substituting $\Lambda(r_u)$ from Eq.~\eqref{eq:intensity}, we can complete the proof. \qedhere
\end{proof}
\end{theorem}

\textbf{Remark} \textit{(GU Location Dependency and CN Density Impact)}: From Theorem~\ref{theorem:success}, we observe that for a given UAV with altitude \( h \), the success probability \( P_{\text{success}}(r_u) \) decreases as the GU's distance \( r_u \) from the UAV increases. This decrease occurs because GUs farther from the UAV experience longer offloading delays (\( t_1 \propto \log(r_u^2 + h^2) \)), which reduces the residual time budget available for forwarding (\( t_2 \)) and computing (\( t_c \)).
Moreover, for a given GU location, a higher CN density \( \lambda_c \) increases the spatial availability of CNs, thereby improving the task completion probability. Specifically, the task completion probability improves exponentially with the void probability obtained from  PPP, given by \( \exp(-\Lambda(r_u)) \).

To evaluate system-level performance, we must consider the spatial distribution of GUs. The overall task completion probability, accounting for these factors, is provided in the following result (Theorem~\ref{theorem:average_probability}).

\begin{theorem}\label{theorem:average_probability}
The spatially averaged task completion probability for GUs uniformly distributed within the~\textit{request zone} (with radius \( R_u \)) is given by: 
\begin{equation}
\overline{P}_{\text{success}} = \frac{2}{R_u^2}\int_0^{R_u} P_{\text{success}}(r_u)\cdot r_u \, dr_u,
\end{equation}
where \( P_{\text{success}}(r_u) \) is defined in Eq.~\eqref{eq:success}.   
\begin{proof} 
Owing to the uniform spatial distribution of GUs, the radial distance \( r_u \) follows the probability density function (PDF):  
\begin{equation}  
f_{r_u}(r_u) = \frac{2r_u}{R_u^2}, \quad 0 \leq r_u \leq R_u.  
\end{equation}  
By the law of total probability, the system-wide task completion probability equals the expectation of \( P_{\text{success}}(r_u) \) over this distribution:  
\begin{equation}  
\overline{P}_{\text{success}} = \mathbb{E}_{r_u}\left[ P_{\text{success}}(r_u) \right] = \int_0^{R_u} P_{\text{success}}(r_u) f_{r_u}(r_u) \, dr_u.  
\end{equation}  
Substituting \( f_{r_u}(r_u) \), we can complete the proof.    
\quad \qedhere
\end{proof}
\end{theorem}

\textbf{Remark} \textit{(Analytical Intractability)}: The nested integrals in $\overline{P}_{\text{success}}$ cause the difficulty in obtaining closed-form solutions due to three fundamental challenges. First, the coverage radius $r_c^{\mathrm{max}}(r_u)$ of the communication-effective~\textit{service zone} exhibits implicit nonlinear dependence on $r_u$ via the latency constraint (Eq. \eqref{eq:max_coverage}), coupling with the integration domains of GU and CN locations. This interdependency prevents decoupling into separable integrals over $r_u$ and $r_c$. Second, the residual time budget $T_{\text{res}}=T_{\text{max}} - t_1(r_u) - t_2(r_c)$ within the integrand $F_{t_c}(T_{\text{res}}; D)$ inherits complexity from both components: $t_1(r_u)$ contains a logarithmic term $\log(r_u^2 + h^2)$, while $t_2(r_c)$, though ostensibly a function of $r_c$, indirectly depends on $r_u$ through the latency-constrained integration upper limit $r_c^{\mathrm{max}}(r_u)$. Third, the computing latency CDF $F_{t_c}(T_{\text{res}}; D)$ itself could be non-trivial. For instance, if $t_c$ includes the queueing delays in stochastic computing systems, its CDF may lack explicit analytical closed forms. These intertwined nonlinearities in spatial, temporal, and statistical dimensions lead us to resort to numerical integration techniques or stochastic geometry-based approximations for practical evaluation. Despite this analytical intractability in terms of closed-form expressions, the integrals in $P_{\text{success}}(r_u)$ and $\overline{P}_{\text{success}}$ remain low-dimensional and can be efficiently evaluated using standard numerical quadrature, which is the approach adopted in our performance evaluation. Moreover, when more general air–ground channels with small-scale fading are considered, the proposed framework can be extended by introducing random channel gains into the received-power expressions and taking expectations over their distributions. This introduces at most one additional integration dimension but preserves the overall semi-analytical structure and the order of computational complexity of the task completion probability evaluation.

\section{Maximizing Task Completion under Energy Constraints}
In this section, we investigate the performance optimization of the UAV-CPNs enabled by hybrid fuel cell and battery-powered UAVs. These hybrid energy architectures, whether implemented in serial, parallel, or decoupled configurations, exhibit distinct power delivery dynamics. However, they share a critical operational limitation: task computing failure can occur if either the battery energy is exhausted or the fuel supply is depleted. This highlights the necessity of joint energy-aware optimization to effectively balance communication and propulsion energy demands under heterogeneous energy sources.

To maximize the task completion probability under dual energy constraints, we develop a systematic optimization framework that explicitly accounts for the coupling between UAV transmit power ($P_d$) and operational altitude ($h$). First, we establish a comprehensive energy consumption model that captures both communication energy and propulsion energy, both of which are functions of $P_d$ and $h$. Then, we formulate a joint optimization problem, aiming to maximize the average task completion probability, subject to realistic energy, hardware, and regulatory constraints.
Finally, we propose an efficient alternating iterative optimization framework that enables real-time adaptation of transmit power and altitude, ensuring reliable and energy-efficient operation in dynamic UAV-CPN environments.

\subsection{Energy Consumption Modeling}
In hybrid fuel cell and battery-powered UAV-CPNs, the energy consumption of the UAV arises from two parts: propulsion (including both mobility and hovering) and communication energy. Since this paper mainly focuses on the vertical movement, we establish an energy model that quantifies propulsion energy consumption as function of UAV altitude \( h \) and transmit power \( P_d \), expressed as~\cite{zhang2020energy}:
\begin{equation}\label{eq:energy_prop}
\begin{split}
E_{\text{prop}}(P_d, h) = &  
\left[(1{+}c)\frac{W^{3/2}}{\sqrt{2\rho A}} + \frac{\delta \rho S_{\text{blade}} v_{\text{tip}}^3}{8}\right] \cdot T(P_d, h)\\&+G(h-h_0),
\end{split}
\end{equation}
where $h_0$ is the initial altitude of the UAV, $T(P_d, h) = t_1(P_d, h)+t_2(P_d, h)$ (refer to Eqs.~\eqref{eq:t1} and~\eqref{eq:t2}) represents hovering duration for \textit{GU-to-UAV task offloading} and \textit{UAV-to-CN forwarding},
\( G \) is the UAV's weight in Newtons (N), \( \rho \) denotes the air density in kilograms per cubic meter (kg/m\(^3\)), \( A \) corresponds to the rotor disc area in square meters (m\(^2\)), \( c \) serves as the induced power correction factor accounting for non-ideal aerodynamic effects, \( \delta \) quantifies the profile drag coefficient of rotor blades, \( S_{\text{blade}} \) indicates the total blade area in square meters (m\(^2\)), and \( v_{\text{tip}} \) specifies the blade tip speed in meters per second (m/s).

In UAV-CPNs, communication energy consumption refers to the energy consumed by transmitting tasks from the UAV to CNs, given by:  
\begin{equation}\label{eq:energy_comm}
E_{\text{comm}}(P_d, h) = P_d \cdot t_2(P_d, h), 
\end{equation}
where $t_2(P_d, h)$ denotes transmission time for \textit{UAV-to-CN forwarding} according to Eq.~\eqref{eq:t2}.

\subsection{Energy-aware Optimization Problem Formulation}
To ensure broad applicability across diverse hybrid UAV-CPN architectures, we develop a generalized modeling framework in which communication energy consumption is constrained by the available battery capacity ($E_{\text{battery}}$), while propulsion energy consumption is limited by the fuel budget ($E_{\text{fuel}}$). Although certain simplifying assumptions are made for analytical tractability, this decoupled-yet-interdependent representation enables a systematic analysis of the complex interplay among energy usage, UAV transmit power ($P_d$), and operational altitude ($h$).  

As established in Section IV, UAV's transmit power and altitude jointly determine CN coverage, which in turn influences critical parameters such as transmission distance, channel conditions, and CN capabilities. These factors collectively govern task computing latency, thereby dictating the UAV’s required hovering duration. Moreover, communication energy consumption is directly impacted by the transmit power itself, creating a dual dependency on both transmit power and altitude.  

Operational altitude further affects mobility-related energy consumption through its influence on mobility distance. Beyond its role in UAV-to-CN communication and computing, altitude also impacts UAV-to-GN transmissions. Prolonged task computing latency increases hovering time, thereby elevating propulsion energy demands. Consequently, both mobility-related and hovering energy consumption contribute to the total propulsion energy burden.

Thus, we formulate a constrained optimization problem that simultaneously addresses: i) communication energy consumption bound by battery capacity ($E_{\text{bat}}$), ii) propulsion energy consumption limited by fuel budget ($E_{\text{fuel}}$), iii) transmit power confined within hardware specifications ($[P_{\min}, P_{\max}]$), and iv) operational altitude restricted by airspace safety protocols ($[h_{\min}, h_{\max}]$). The joint optimization problem can be mathematically formulated as:
\begin{subequations}
\begin{align}
(\textbf{P1})\,\max_{P_d,h} \quad & \overline{P}_{\text{success}}(P_d, h) \\
\text{s.t.} \quad & E_{\text{comm}}(P_d, h) \leq E_{\text{battery}}, \label{eq:constraint1} \\
& E_{\text{prop}}(P_d, h) \leq E_{\text{fuel}}, \label{eq:constraint2} \\
& P_d \in [P_{\min}, P_{\max}], \label{eq:constraint3} \\
& h \in [h_{\min}, h_{\max}], \label{eq:constraint4}
\end{align}
\end{subequations}
where $\overline{P}_{\text{success}}(P_d, h)$ denotes the average task completion probability derived from analysis in Section~IV,~\eqref{eq:constraint1} and~\eqref{eq:constraint2} represent the battery and fuel energy budgets, respectively, and~\eqref{eq:constraint3} and~\eqref{eq:constraint4} define the transmit power constraint and the operational altitude range, respectively. 

This formulation exhibits two characteristics that challenge conventional optimization methods. First, the altitude-dependent LoS probability $P_{\text{LoS,up}}$ introduces non-linear coupling between $h$ and $P_d$.
Second, the dual-energy constraints create discontinuous feasible regions in the parameter space. To address~\textbf{P1} while maintaining computational efficiency for real-time implementation, we develop an alternating iterative joint optimization strategy.

\begin{algorithm}[t]
\caption{Alternating Iterative Joint Optimization}
\label{alg:alternating_optim}
\SetAlgoLined
\DontPrintSemicolon
\KwIn{Energy budget ($E_{\text{battery}}, E_{\text{fuel}}$)}
\KwOut{Optimized parameters $(h^*, P_d^*)$.}
\textbf{Initialization:}\;
Initial altitude $h^{(0)}$, initial transmit power $P_d^{(0)}$, maximum iterations~$\textit{max\_iter}$, convergence threshold~$\epsilon$, $\eta_{\text{prev}} \gets 0$\;

\For{$k = 1$ \KwTo $\text{max\_iter}$}{
    \textbf{Altitude optimization:}\;
    \begin{equation*}
        h^{(k)} = \arg\max_h\overline{P}_{\text{success}}(h, P^{(k-1)}; E_{\text{battery}}, E_{\text{fuel}})
    \end{equation*}
    \textbf{Transmit power optimization:}\;
    \begin{equation*}
        P_d^{(k)} = \arg\max_{P_d} \overline{P}_{\text{success}}(h^{(k)}, P; E_{\text{battery}}, E_{\text{fuel}})
    \end{equation*}

    \textbf{Convergence Check:}\;
    $\eta^{(k)} \gets \overline{P}_{\text{success}}(h^{(k)}, P_d^{(k)})$\;
    
    \If{$|\eta^{(k)} - \eta_{\text{prev}}| < \epsilon$}{
        \KwRet $(h^{(k)}, P_d^{(k)})$
    }
    $\eta_{\text{prev}} \gets \eta^{(k)}$\;
}
\KwRet $(h^{(\text{max\_iter})}, P_d^{(\text{max\_iter})})$
\end{algorithm}

\subsection{Alternating Iterative Joint Optimization Framework}
In this section, we propose a robust iterative method to solve \textbf{P1}, tailored to the practical constraints of hybrid-powered UAV-CPNs under dynamic CN accessibility.

The proposed joint optimization strategy operates through the following iterations. First, we initialize the UAV's altitude \( h^{(0)} \), transmit power \( P_d^{(0)} \), the maximum iteration count \( \textit{max\_iter} \), and convergence threshold \( \epsilon \). Subsequently, we employ a two-stage iterative optimization approach: While maintaining fixed transmit power, the altitude is optimized through golden-section search to maximize the task completion probability. This derivative-free approach effectively handles non-monotonic relationships between altitude and system performance metrics (e.g., task completion probability and energy consumption). Following this altitude optimization, we employ a modified quasi-Newton method combining gradient descent with approximate Hessian information to optimize transmit power parameters. Adaptive learning rate decay will be adopted to prevent oscillations during gradient direction changes. These alternating updates continue until the relative improvement in the task completion probability over a sliding window of the last few iterations falls below \( \epsilon \), or \( \textit{max\_iter} \) is reached, providing enhanced convergence stability. Upon convergence, the solution is validated against both communication and propulsion energy constraints. If either constraint is violated, the system automatically reverts to a predefined safe operating point (e.g., \( h = 50 \)\,m, \( P_d = 10 \)\,dBW), ensuring operational feasibility under model uncertainty or environmental variations. These alternating updates continue until the relative improvement in the task completion probability over a sliding window of the last \( W \) iterations (e.g., \( W = 5 \)) falls below \( \epsilon \), or \( \textit{max\_iter} \) is reached, providing enhanced convergence stability. Upon convergence, the solution is validated against both communication and propulsion energy constraints. If either constraint is again violated, the system automatically reverts to a predefined safe operating point (e.g., \( h = 50 \)\,m, \( P_d = 10 \)\,W), ensuring operational feasibility under model uncertainty or environmental variations.

The time complexity of this algorithm is determined by three main factors: i) the per-iteration computational cost of altitude and transmit power optimization, ii) the number of iterations required for convergence, and iii) the user population size \( N \). At each iteration, altitude optimization via the golden-section method involves a fixed number of steps, where each step evaluates the objective function, resulting in a cost of \( \mathcal{O}(n_{\text{steps\_h}} \cdot N) \). Similarly, optimizing transmit power through the modified quasi-Newton method requires a fixed number of steps, each step computing gradients and Hessians information of the objective function, leading to \( \mathcal{O}(n_{\text{steps\_p}} \cdot N) \). Assuming a maximum of \( T \) iterations, the total time complexity becomes \( \mathcal{O}(T \cdot (n_{\text{steps\_h}} + n_{\text{steps\_p}}) \cdot N) \). Since \( n_{\text{steps\_h}} \) and \( n_{\text{steps\_p}} \) are constants, this is simplified to \( \mathcal{O}(T \cdot N) \). The dominant factor is the linear dependence on \( N \), as evaluating the objective function requires iterating over all users to compute channel states and energy constraints. In practice, convergence often occurs far earlier than the theoretical maximum \( T \), making the method scalable for large networks~\cite{zhang2020energy,Pervez2024energy}.

\section{Numerical results}
In this section, we evaluate the task completion probability using parameters listed in Tables~\ref{tab:parameter_network} and~\ref{table:parameter_uav} unless explicitly indicated. These parameters align with the scenarios where UAVs and GUs employ wireless transmission modules to support low-latency services like real-time video analytics~\cite{Fang2025}. Our evaluation focuses on four critical aspects: i) the validation of theoretical models, ii) the trade-off among communication-computing resources through strategic positioning of UAV altitude, iii) the potential for performance enhancement with more reachable CNs, and iv) energy-constrained performance optimization and parameter sensitivity analysis. 

\begin{table}[t]\label{table:parameter_network}
    \centering
    \caption{Parameter setting for CPNs}
    \label{tab:parameter_network}
    \resizebox{\linewidth}{!}{
    \begin{tabular}{ccc}
        \toprule
        \textbf{Description} & \textbf{Parameter} & \textbf{Value} \\
        \midrule
        Transmit power of GUs &  $P_u$ & $20$ dBW\\
        Path loss exponent & $\alpha_u$ & $2$ \\
        NLoS attenuation coefficient & $\eta$ & $20$dB \\
        Bandwidth & $W$ & 8 MHz \\
        Noise power & $N_0$ & $-120$ dBm \\
        Data size & $D$ & 1 MB \\
        Maximum allowable latency & $T_{\text{max}}$ & $55$ ms \\
        Node density (GUs, CNs) & $\lambda_u$, $\lambda_c$ & $500$, $5$ nodes/km$^2$ \\
        Zone radius (Request, service) & $R_\text{u}$, $R_\text{d}$& $200$, $1000$ m \\
        Parameters for urban environment & $B$, $C$ & $0.136$, $11.95$\\
        \bottomrule
    \end{tabular}}
\end{table}

\begin{table}[t]
    \centering
    \caption{Parameter setting for the UAV}
    \label{table:parameter_uav}
    \resizebox{\linewidth}{!}{
    \begin{tabular}{ccc}
        \toprule
        \textbf{Description} & \textbf{Parameter} & \textbf{Value} \\
        \midrule
        Incremental correction factor in~\eqref{eq:energy_prop} &  $c$ & $0.1$  \\
        UAV take-off mass & $G$ & $26$ kg \\
        Air density & $\rho$ & $1.225 kg/m^3$ \\
        Rotor disc area & $A$ & $1 m^2$ \\
        Profile drag coefficient & $\delta$ & $0.012$ \\
        Total blade area & $S_{\text{blade}}$ & $0.2 m^2$ \\
        Speed of rotor blade tip & $v_{\text{tip}}$ & $250$ m/s \\
        Battery energy budget & $E_{\text{battery}}$ & $20-120$ J \\
        Fuel cell energy budget & $E_{\text{fuel}}$ & $3\times10^4-6\times10^4$ J \\
        UAV initial altitude & $h_0$ & $50$ m \\
        Transmit power of UAV &  $P_d$ & $20$ dBW \\
        \bottomrule
    \end{tabular}}
\end{table}

\subsection{Evaluation on task completion probability}
\begin{figure}[b]
    \centering
    \includegraphics[width=0.825\linewidth]{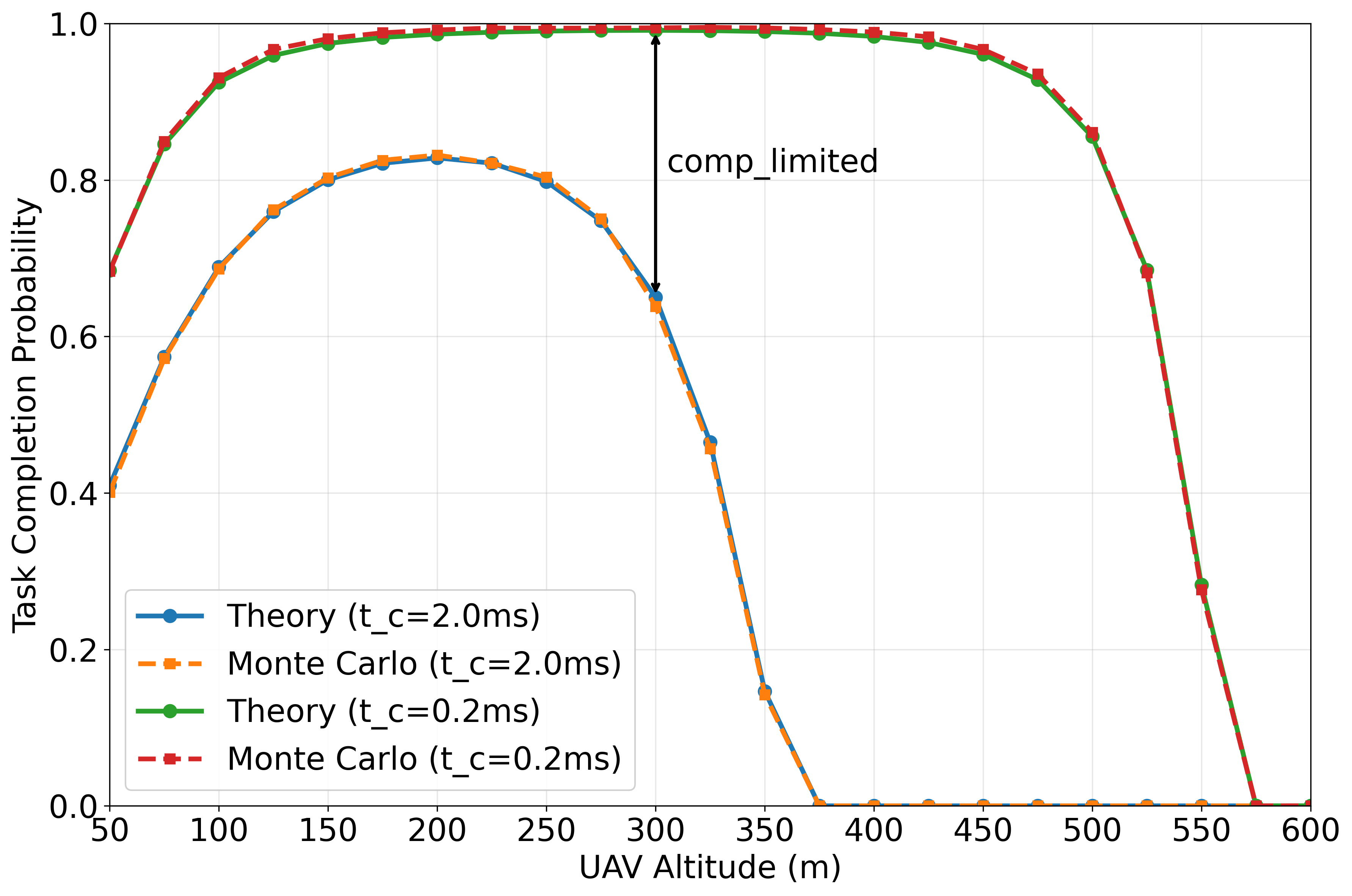}
    \caption{Task completion probability vs. UAV altitude.}
    \label{fig:success-probability-analysis}
\end{figure}

We first validate the analytical expressions derived in Theorem~\ref{theorem:average_probability} through extensive Monte Carlo simulations (10,000 runs), each involving random sampling of $400$ GUs for average task completion probability calculation. As shown in Fig.~\ref{fig:success-probability-analysis}, our theoretical analysis (\textit{Theory} curves) exhibits close alignment with Monte Carlo simulation results (\textit{Monte Carlo} curves) across two representative computing capability scenarios: $t_c = 0.2$\,\text{ms} and $t_c = 2$\,\text{ms}, respectively. System performance degrades at both extremely low and high UAV altitudes.
Specifically, low altitudes result in NLoS-dominated link connections between the UAV and GUs or CNs, while high altitudes lead to excessive path loss, both of which restrict the communication coverage of the UAV and thus fundamentally limit the effective utilization of distributed computing power. Numerical results confirm the existence of an optimal UAV altitude (approximately $200$\,\text{m} when $t_c = 2$\,\text{ms}) that resolves communication bottlenecks through adaptive UAV positioning. Moreover, enhanced computing power (reducing $t_c$ from $2$\,\text{ms} to $0.2$\,\text{ms}) mitigates computing bottlenecks (marked as \textit{comp\_limited}), yielding altitude-dependent performance gains. The UAV-CPN architecture consequently necessitates a dynamic configuration that involves both communication and computing simultaneously.

\begin{figure*}[t] 
    \centering
    \begin{minipage}[ht]{0.475\textwidth}
        \centering
        \includegraphics[width=\linewidth]{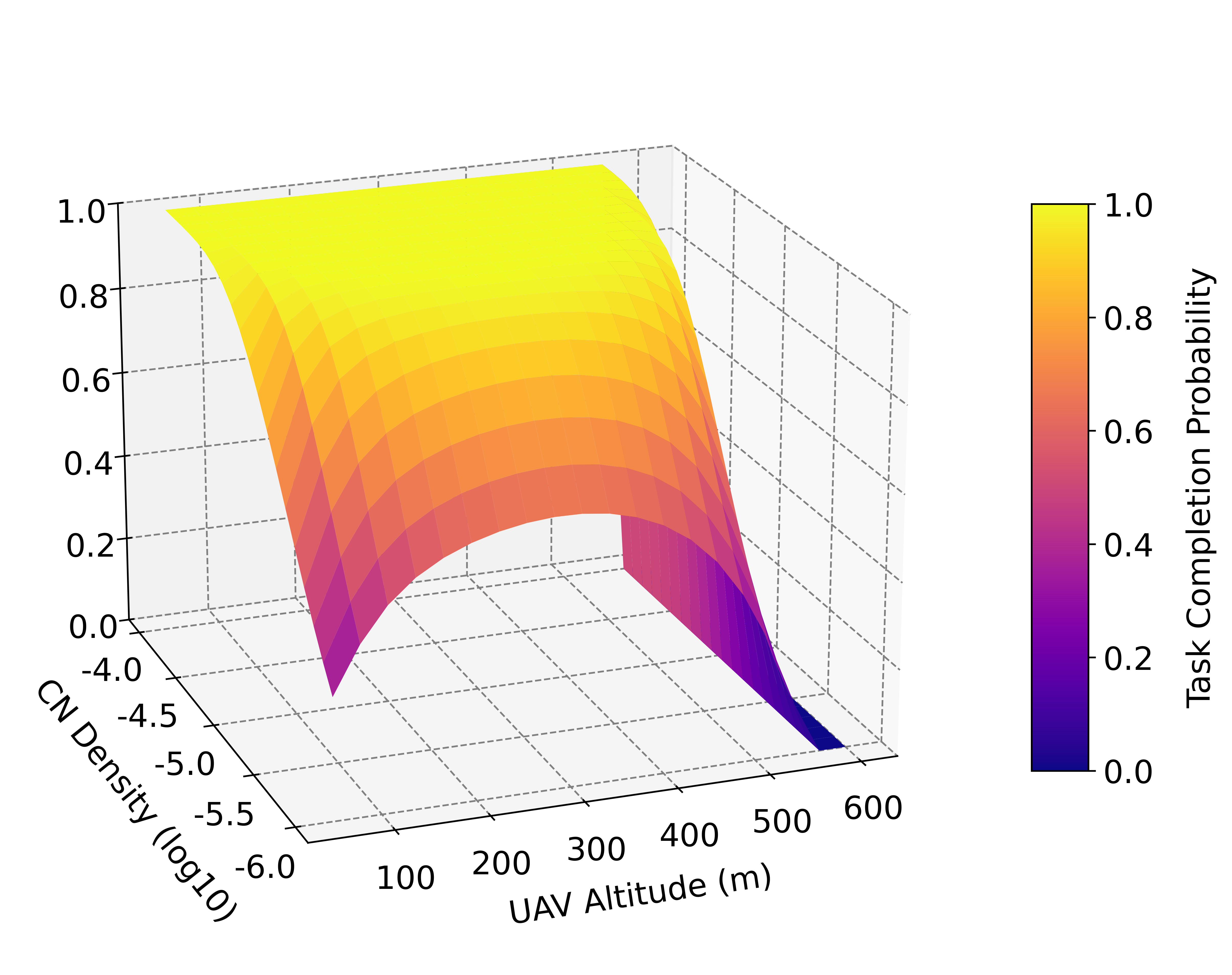}
        \caption{Task completion probability vs. CN density \& UAV altitude.}
        \label{fig:3d_altitude_density_success}
    \end{minipage}
    \hfill 
    \begin{minipage}[ht]{0.475\textwidth}
        \centering
        \includegraphics[width=\linewidth]{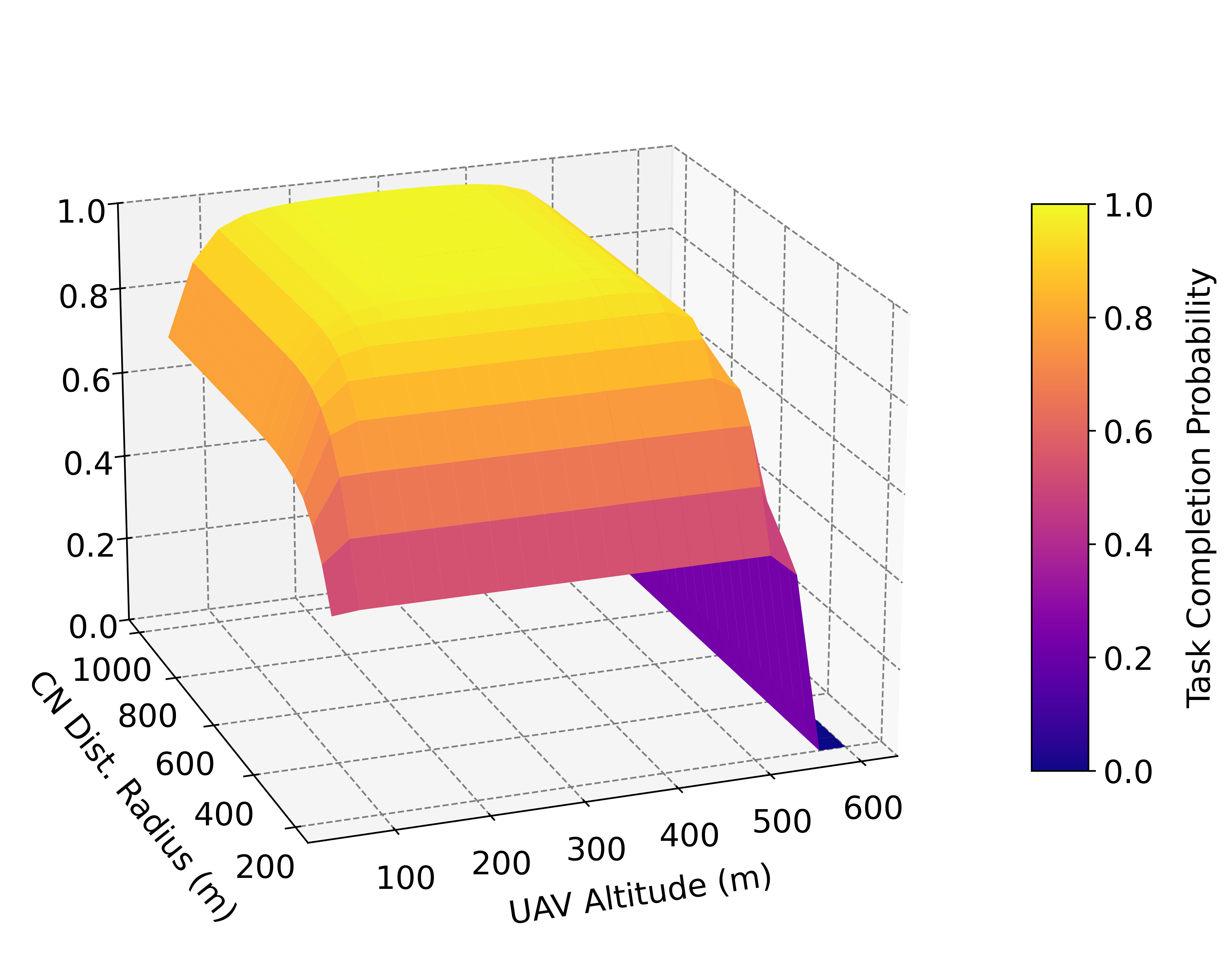}
        \caption{Task completion probability vs. CN distribution \& UAV altitude.}
        \label{fig:3d_altitude_radius_success}
    \end{minipage}
\end{figure*}

\begin{figure*}[ht]
    \centering
    \includegraphics[width=0.825\linewidth]{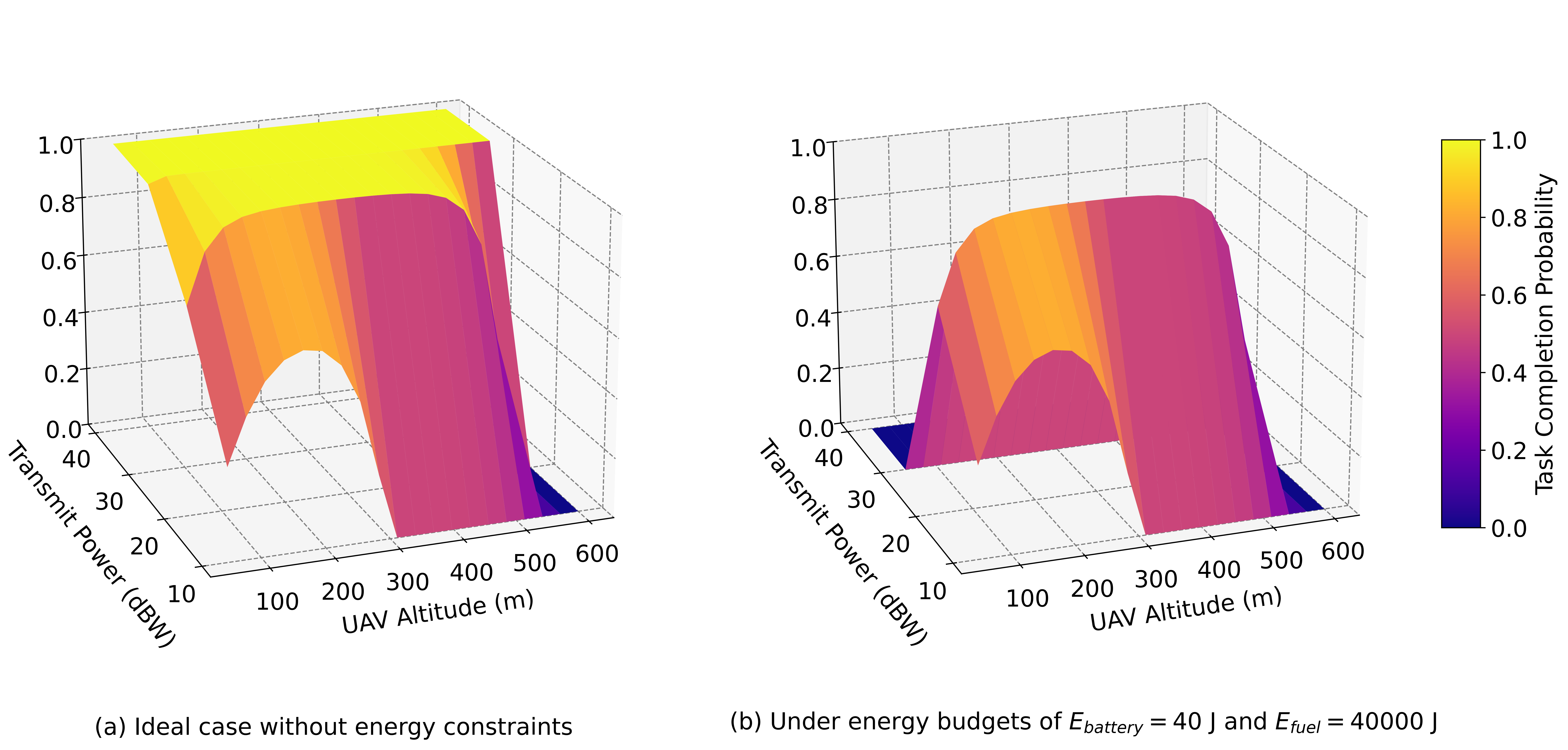}
    \caption{The joint impact of UAV transmit power and altitude on task completion probability under different energy supply conditions: (a) Ideal case without energy constraints, (b) Under energy budgets of $E_{\text{battery}}=40$\,\text{J} and $E_{\text{fuel}}=40,000$\,\text{J}.}
    \label{fig:3d_comparison}
\end{figure*}

We further investigate the fundamental trade-off between communication coverage and computing power by analyzing the joint effects of CN density and UAV altitude on task completion probability. Fig.~\ref{fig:3d_altitude_density_success} presents a $3$D surface plot that reveals a nonlinear interdependency between these two critical parameters and highlights their varying sensitivities. For example, when the CN density is low, the task completion probability is highly sensitive to the altitude of the UAV. Therefore, careful selection of the UAV altitude is crucial to achieve a high task completion probability under such conditions. In contrast, when the CN density is high, the altitude of the UAV can vary in a wide range while still maintaining a high task completion probability.

\begin{figure*}[t!]
    \centering
    \includegraphics[width=0.75\textwidth]{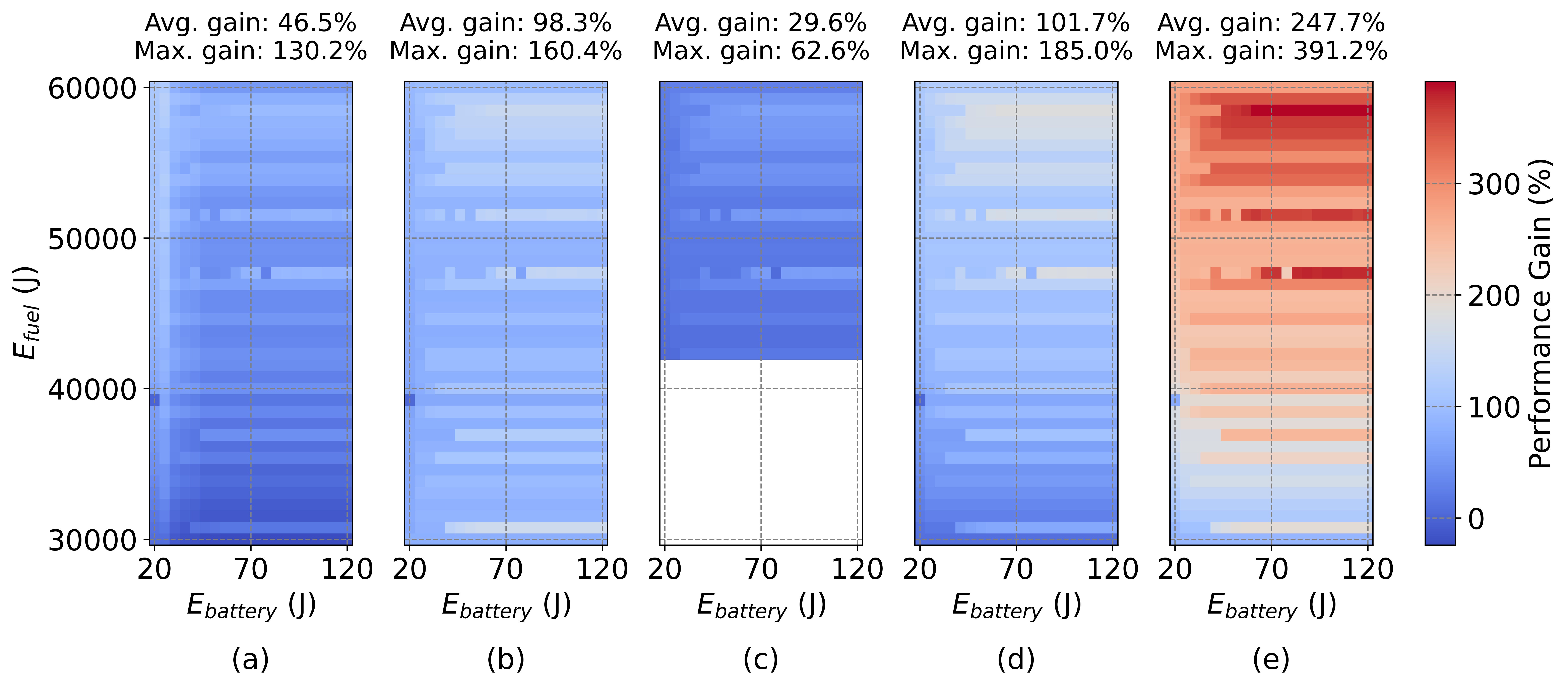}
    \caption{Performance gain of the proposed joint optimization (altitude and transmit power) over key baseline strategies under varying energy budget combinations: (a) Transmit Power Only ($h =50$\,\text{m}), (b) Altitude Only ($P_d=5$\,\text{dBW}), (c) Static Configuration ($h =100$\,\text{m}, $P_d=10$\,\text{dBW}), (d) Static Configuration ($h =50$\,\text{m}, $P_d=10$\,\text{dBW}), (e) Static Configuration ($h =50$\,\text{m}, $P_d=5$\,\text{dBW}). The proposed method dynamically balances coverage, link quality, and energy constraints, achieving up to 29.6\% higher task completion probability than the best-performing baseline. This highlights the critical importance of co-optimizing UAV altitude and transmit power in UAV-CPNs. Notably, while the static configuration with $h = 100$\,\text{m} and $P_d = 10$\,\text{dBW} occasionally achieves high performance, its practical feasibility is severely limited--40\% of energy budget combinations are infeasible (indicated by white regions in Fig.~6c), resulting in complete task failure.}
    \label{fig:gain_strategy_heatmaps}
\end{figure*}

Next, we investigate the performance gain achieved through CN distribution radius expansion. Fig.~\ref{fig:3d_altitude_radius_success} shows a 3D surface plot that reveals the coupled effects of CN spatial distribution and UAV altitude on task completion probability.
For fixed UAV altitudes (e.g., at a UAV altitude of $300$m), extending CN distribution beyond the~\textit{request zone} boundaries (specifically when the distribution radius exceeds $200$\text{m}) produces substantial performance gains. Specifically, expanding the CN distribution radius from $200$\text{m} (corresponding to our prior work~\cite{Deng2024})  to $1,000$\text{m} achieves a $2.13\times$ improvement in task completion probability ($46.65$\% $\rightarrow$ $99.14$\%), confirming our framework's capability to leverage \textit{ubiquitous computing power distribution}. For fixed CN distributions, we confirm the existence of an altitude-dependent performance maximum, demonstrating the effectiveness of~\textit{dynamic CN accessibility control} , which is consistent with our prior work~\cite{Deng2024}. While the UAV-CPN architecture introduces multi-dimensional coordination complexity, this proves to be essential for overcoming individual communication or computing bottlenecks, ultimately achieving superior performance.

Finally, we evaluate the joint impact of UAV transmit power and altitude on task completion probability under different energy budgets. As shown in Fig.~\ref{fig:3d_comparison}, our 3D surface analysis reveals a significant energy-performance tradeoff in UAV-CPNs. The constrained scenario (Fig.~\ref{fig:3d_comparison}b) demonstrates severe performance degradation at high power ($\ge30$\,\text{dBW}) and high altitude ($\ge500$\,\text{m}) compared to the ideal case (Fig.~\ref{fig:3d_comparison}a). For example, at a configuration of $30$\,\text{dBW}/$310$\,\text{m}, the task completion probability drops from $1$ under ideal conditions to nearly $0$ under energy constraints due to excessive communication energy consumption. On average, the drop across different energy budgets is approximately $50$\%.
This analysis confirms that energy constraints create critical performance bottlenecks, necessitating energy-aware UAV deployment strategies in UAV-CPNs. In the following section, we systematically evaluate optimization approaches to maintain service quality under strict energy budget constraints.

\subsection{Energy-Constrained Performance Analysis}
We evaluate the performance gain achieved by jointly optimizing the UAV altitude and transmit power (hereafter referred to as the~\textit{joint optimization} strategy) under predefined battery ($E_{\text{battery}}$) and fuel ($E_{\text{fuel}}$) energy budgets, compared to three types of baseline configurations.
\begin{figure*}[t!]
    \centering
    \includegraphics[width=0.75\textwidth]{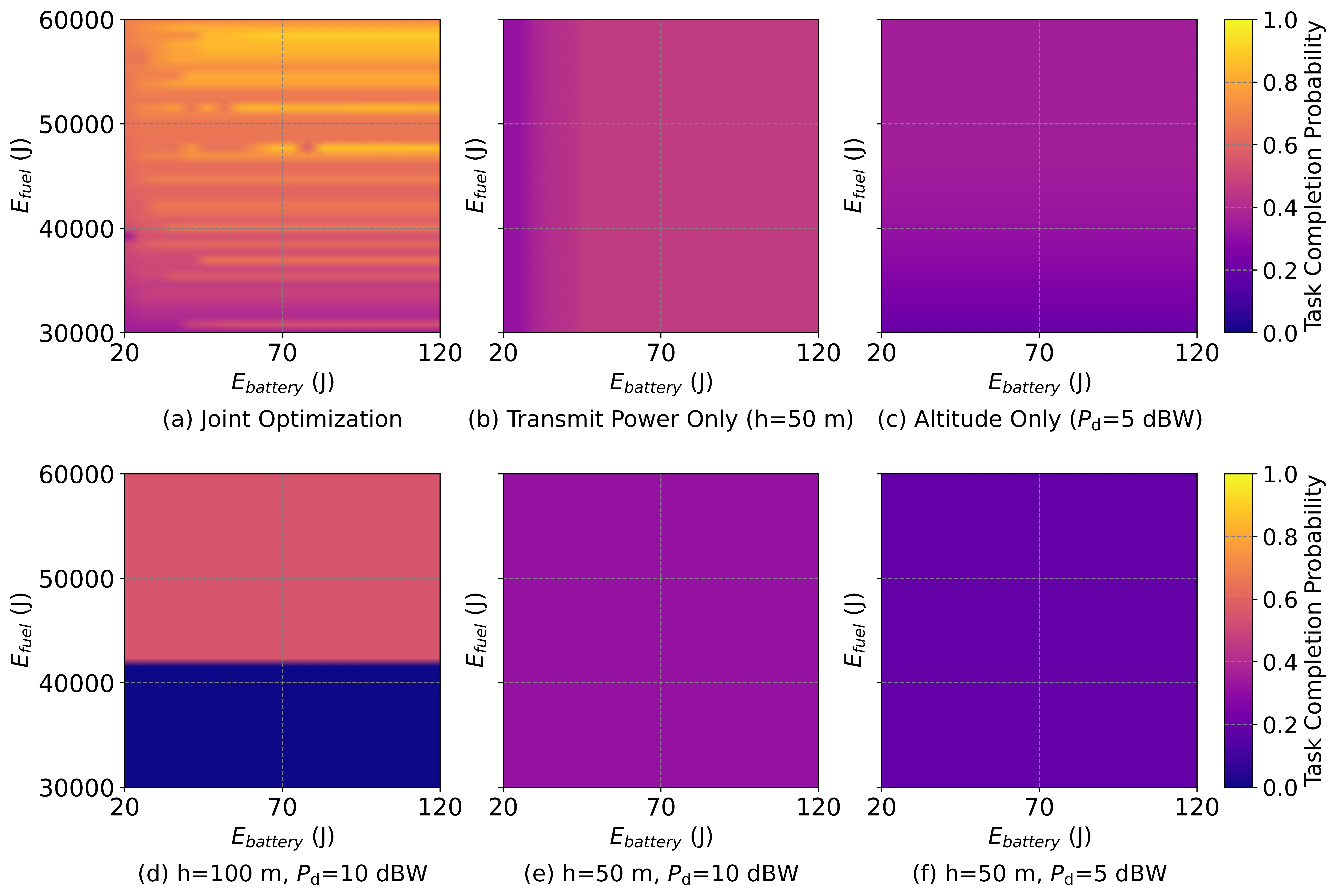}
    \caption{Task completion probability under varying energy budget combinations: (a) Joint Optimization, (b) Transmit Power Only ($h =50$\,\text{m}), (c) Altitude Only ($P_d=5$\,\text{dBW}), (d) Static Configuration ($h =100$\,\text{m}, $P_d=10$\,\text{dBW}), (e) Static Configuration ($h =50$\,\text{m}, $P_d=10$\,\text{dBW}), (f) Static Configuration ($h =50$\,\text{m}, $P_d=5$\,\text{dBW}). The proposed joint optimization achieves the highest task completion probability across all energy budget combinations while baseline strategies suffer from limited flexibility or frequent infeasibility, particularly under stringent energy conditions.}
    \label{fig:multi_strategy_heatmaps}
\end{figure*}

\begin{itemize}  
    \item \textit{Transmit Power Only}: Optimize the UAV's transmit power while fixing the altitude at \( h = 50 \, \text{m} \). This baseline isolates the impact of power allocation.  
    \item \textit{Altitude Only}: Optimize the UAV's altitude while fixing the transmit power at \( P_d = 5 \, \text{dBW} \). This baseline isolates the impact of altitude control.  
    \item \textit{Static Configurations}: Three pairs of static parameters: 
    \begin{itemize}  
        \item High altitude, low power: $h=100$~m, $P_d = 10$~dBW.  
        \item Low altitude, high power: \( h = 50 \, \text{m}, P_d = 10 \, \text{dBW} \).  
        \item Conservative operation: \( h = 50 \, \text{m}, P_d = 5 \, \text{dBW} \).  
    \end{itemize}  
\end{itemize} 
The analysis aims to: i) quantify the benefits of joint optimization, and ii) identify which parameter (altitude or transmit power) has a more pronounced impact on task completion probability under different combination of energy constraints $E_{\text{battery}}$ and $E_{\text{fuel}}$. The performance gain is defined as the relative improvement in task completion probability:  
\begin{equation}  
    \text{Performance Gain (\%)} = \frac{\mathcal{P}_{\text{joint}} - \mathcal{P}_{\text{baseline}}}{\mathcal{P}_{\text{baseline}}} \times 100.  
\end{equation}  
where \( \mathcal{P}_{\text{joint}} \) and \( \mathcal{P}_{\text{baseline}} \) denote the task completion probabilities of the joint optimization and baseline strategies, respectively. 

As shown in Fig.~\ref{fig:gain_strategy_heatmaps}, the proposed \textit{Joint Optimization} strategy demonstrates consistently superior performance, achieving strictly positive average gains ranging from $29.6$\% to $247.7$\%, against five baseline strategies, with peak improvements exceeding 390\%. These results quantitatively validate its capability to maximize task completion probability under dual-energy constraints through parameter coordination. Subplots (a) and (b) of Fig.~\ref{fig:gain_strategy_heatmaps} reveal the critical advantage of joint optimization over single-parameter approaches, achieving $46.5$\% average and $130.2$\% maximum gains against power-only optimization, and more substantial $98.3$\% average and $160.4$\% maximum improvements over altitude-only optimization. These findings highlight the essential role of dynamic coordination between transmit power adjustment and altitude adaptation in maximizing the task completion probability. Moreover, the analysis reveals that transmit power optimization contributes more significantly to performance enhancement than altitude optimization, suggesting that CN spatial coverage improvements through power control may outweigh the benefits of that via altitude adjustments under dual-energy constraints. The most significant gains emerge in comparisons with the static \( h=50\, \text{m}, P_{\mathrm{d}}=5\, \text{dBW} \) configuration, reaching a peak improvement of 391.2\%. This stems primarily from the dual limitations inherent to static strategies: fixed low transmit power (\( P_{\mathrm{d}}=5\, \text{dBW} \)) fundamentally restricts the communication coverage capability of CNs, while suboptimal altitude selection amplifies path loss inefficiencies. 
As shown in subplots (c)-(e) of Fig.~\ref{fig:gain_strategy_heatmaps}, while the static configuration with \(h=100\, \text{m}, P_{\mathrm{d}}=10\, \text{dBW}\) occasionally achieves the best performance among three static configurations, its practical viability is severely compromised by 40\% invalid energy budget combinations (indicated by white spaces in subplots (c) of Fig.~\ref{fig:gain_strategy_heatmaps}), where task completion probability collapses to zero. The reported gains derive from 480 valid energy budget combinations, with the static \( h = 100 \, \text{m} \) strategy exhibiting the worst robustness despite achieving the highest baseline task completion probabilities in valid regions. This phenomenon reinforces the necessity of dynamic parameter coordination to maintain reliable performance across the energy budget combination. 
 
In Fig.~\ref{fig:multi_strategy_heatmaps}, we display the concrete task completion probabilities for six UAV-CPN configurations under varying energy budgets (\( E_{\text{battery}} \in [20, 120] \, \text{J}, E_{\text{fuel}} \in [30, 60] \, \text{kJ} \)). These results not only provide the foundation for the performance gain calculation in Fig.~\ref{fig:gain_strategy_heatmaps} but also offer intuitive insights into the system behavior. The \textit{Joint Optimization} strategy achieves the best task completion probability across all energy combinations, obviously outperforming the other baseline strategies. 
Subplots (a) and (b) in Fig.~\ref{fig:multi_strategy_heatmaps} exhibit sensitivity to battery energy budgets \( E_{\text{battery}} \), reflecting the incorporation of transmit power optimization in both \textit{Joint Optimization} and \textit{Transmit Power Only} strategies. Similarly, subplots (a) and (c) in Fig.~\ref{fig:multi_strategy_heatmaps} reveal a significant dependence on fuel energy \( E_{\text{fuel}} \), highlighting the critical role of propulsion energy in altitude optimization.
At low altitudes (e.g., \( h = 50 \, \text{m} \)), optimizing the transmit power (Fig.~\ref{fig:multi_strategy_heatmaps}~(b) vs.~(e)) enhances task completion probability, demonstrating the dominance of power adaptation in energy-constrained regimes. Under low transmit power (\( P_d = 5 \, \text{dBW} \)), altitude optimization (Fig.~\ref{fig:multi_strategy_heatmaps}~(c) and~(f)) increases the task completion probability, validating altitude tuning as a critical lever for communication coverage enhancement. 
A striking observation in static configurations is that when \( E_{\text{fuel}} \leq 42,000 \, \text{J} \), the strategy with \( h = 100 \, \text{m}\), \(P_d = 10 \, \text{dBW} \) fails completely, resulting in zero task completion probability. This corresponds to the white regions depicted in  Fig.~\ref{fig:gain_strategy_heatmaps}~(c), indicating that with these configurations, the fuel energy \( E_{\text{fuel}} \) becomes the primary performance bottleneck. 

\begin{figure*}[ht]
    \centering
    \includegraphics[width=0.825\textwidth]{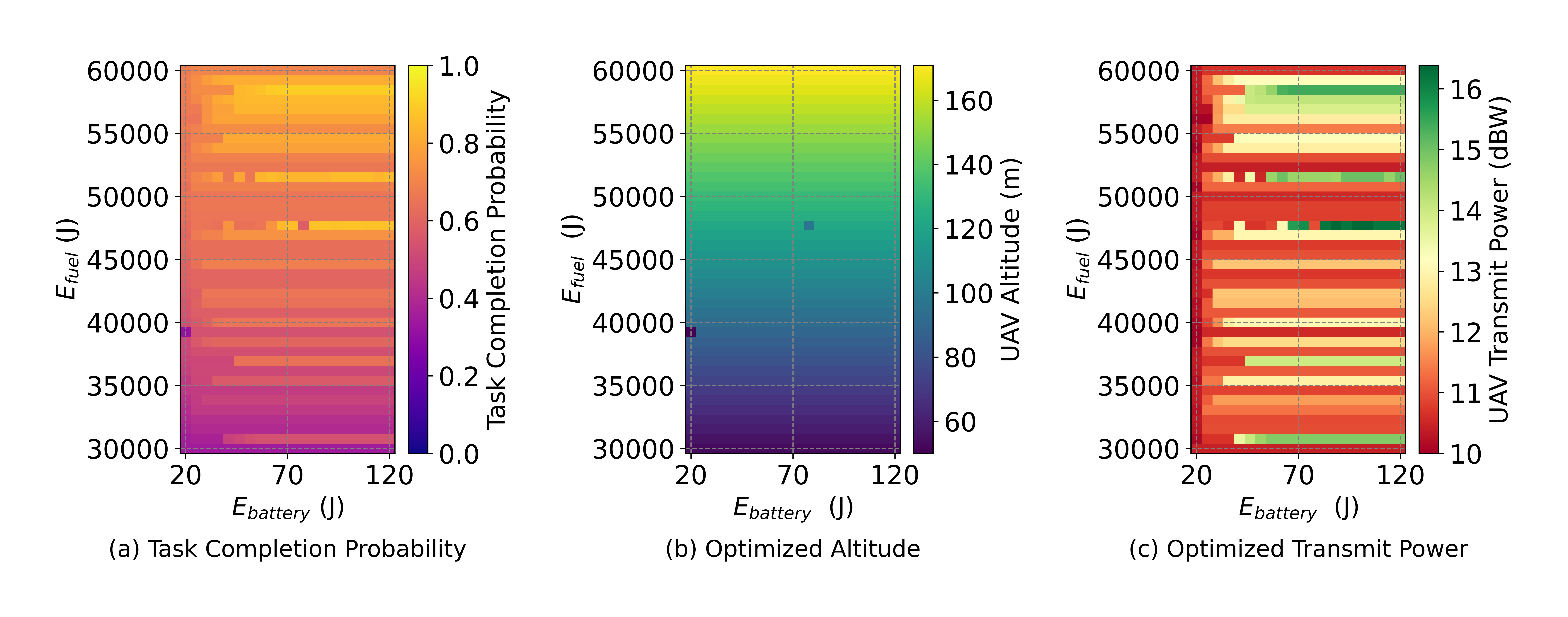}
    \caption{Optimized performance and system parameters under varying battery-fuel cell energy budget combinations: (a) Maximum task completion probability ($P_{\text{success}}^*$), (b) Optimized UAV altitude ($h^*$), and (c) Optimized transmit power ($P_{\text{d}}^*$) obtained via the proposed joint optimization framework.}
    \label{fig:optimized_energy_budget}
\end{figure*}

\begin{figure*}[ht]
    \centering
    \includegraphics[width=0.825\textwidth]{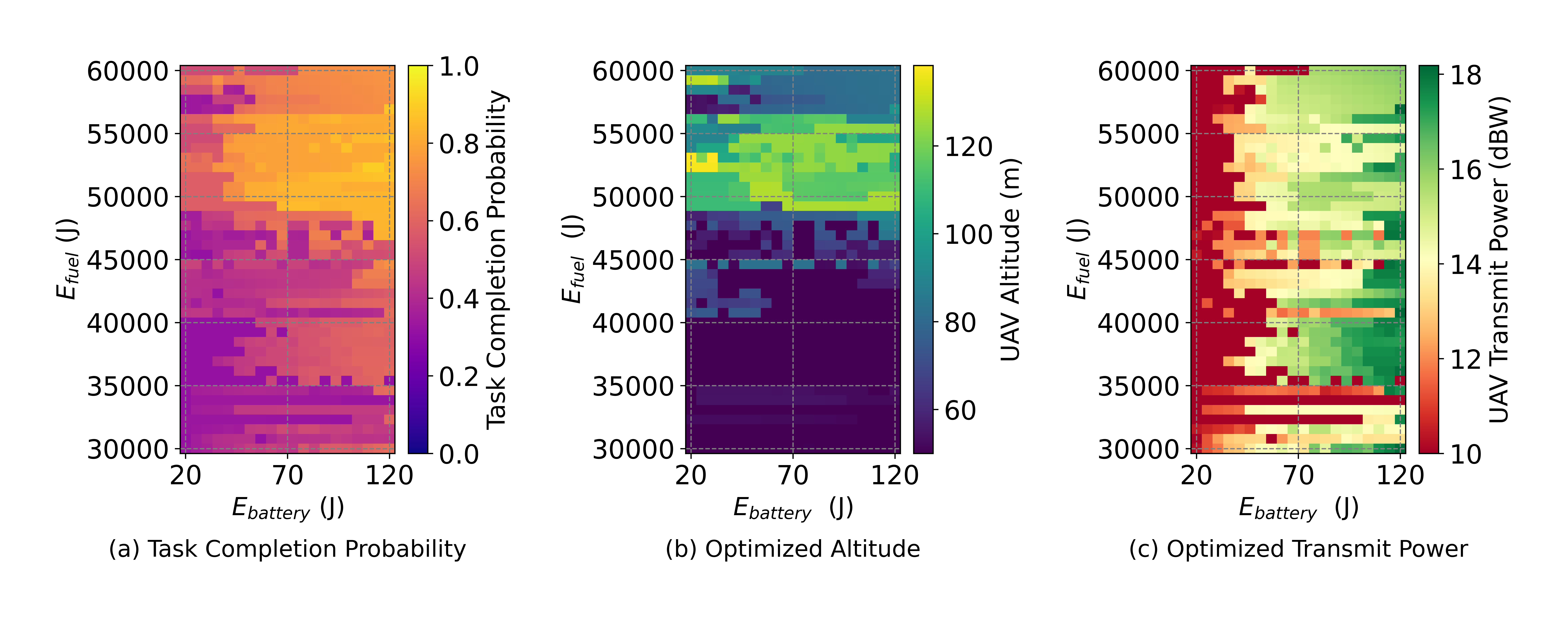}
    \caption{Performance of a Bayesian optimization-based baseline for solving problem P1 under varying battery–fuel cell energy budget combinations: (a) Achieved task completion probability ($P_{\text{success}}^*$), (b) Optimized UAV altitude ($h^*$), and (c) Optimized transmit power ($P_{\text{d}}^*$). This joint optimization approach serves as a comparative benchmark for the proposed two-stage iterative method.}
    \label{fig:optimized_energy_budget_beyesian}
\end{figure*}

The joint optimization of UAV altitude and transmit power under dual energy constraints, specifically from fuel cells and batteries, has not been previously addressed in the literature. Consequently, there are no established baseline mechanisms for direct comparison. 
To evaluate the effectiveness of the proposed alternating iterative joint optimization framework, we compare it against a Bayesian optimization (BO)-based baseline method, which is commonly used for black-box optimization of complex, multi-parameter problems. The results are presented in Figs.~8 and~9.

Subplots (b) and (c) in Fig.~\ref{fig:optimized_energy_budget} illustrate the optimized altitude~(\( h^* \)) and transmit power~(\( P_{\text{d}}^* \)), respectively, obtained from the proposed method, while subplot (a) shows the corresponding task completion probability ($P_{\text{success}}^*$). From these three subplots, we observe that the optimal task completion probability, the optimized UAV altitude, and the optimized transmit power are all sensitive to the fuel energy budget \( E_{\text{fuel}} \). Notably, transmit power exhibits greater sensitivity to both $E_{\text{fuel}}$ and the battery energy budget $E_{\text{battery}}$ than what altitude does. As a result, the spatial pattern of $P_{\text{success}}^*$ in Fig.~8(a) closely resembles that of $P_{\text{d}}^*$ in Fig.~8(c). This suggests that, while both parameters are important, transmit power tuning plays a more dominant role in shaping performance under the tested energy configurations, consistent with the insights from  Fig.~\ref{fig:gain_strategy_heatmaps} and Fig.~\ref{fig:multi_strategy_heatmaps}. The observed parameter sensitivities provide valuable theoretical guidance for energy management in hybrid fuel and battery-powered UAV-CPNs. For instance, when $E_{\text{fuel}} > 50,000$\,J, converting excess fuel energy into electrical energy (e.g., via onboard generators) can alleviate battery energy limitations and thereby improve task completion probability, as demonstrated in Fig.~\ref{fig:optimized_energy_budget}a. This confirms that $E_{\text{battery}}$ can act as a performance bottleneck, aligning with findings in Fig.~\ref{fig:3d_comparison}.

To comprehensively evaluate the proposed optimization methodology, we implement a comparative BO framework~\cite{wang2023recent} for solving the joint parameter optimization problem \textbf{P1}. While Bayesian methods demonstrate established competence in navigating multi-dimensional spaces through Gaussian process-based surrogate modeling, our controlled experimental analysis reveals critical performance limitations due to energy constraints. Under equivalent computational budgets ($30$ function evaluations per energy budget configuration to ensure equitable comparison conditions), the proposed optimization algorithm demonstrates consistent superiority with $13.8$\% average improvement in task completion probability across all energy budgets, achieving peak enhancements of $49.16$\% at critical operational points ($E_{\text{battery}} = 46.32$\,\text{J}, $E_{\text{fuel}} = 58,461.54$\,\text{J}). Spatial performance analysis confirms that our proposed method outperforms BO in $69.2$\% of operational scenarios. As visualized in  Figs.~\ref{fig:optimized_energy_budget}(a) and \ref{fig:optimized_energy_budget_beyesian}(a), it becomes evident that our proposed algorithm achieves superior task completion probabilities. This performance gap primarily stems from: 1)~\textit{The curse of dimensionality:} alternating optimization decomposes the $2$D problem into sequential $1$D optimizations, whereas Bayesian optimization directly searches the full parameter space, resulting in significantly reduced sampling density under equivalent evaluation budgets; 2)~\textit{Constraint-handling efficacy:} our method incorporates explicit feasibility checks during each single-parameter optimization phase, while the Bayesian approach's penalty-based constraint relaxation may induce convergence to infeasible regions. These comparative results validate that our proposed strategy has better suitability for real-world deployment in energy-constrained UAV-CPN systems.  

\section{Conclusion}
In this paper, we have investigated the task completion probability (i.e., task completion rate or task throughput) in UAV-enabled computing power networks, where an aerial UAV relay facilitates the transmission of ground users' computing tasks to distributed computing nodes (CNs) for real-time processing and computing. Our proposed framework enables tasks generated from a constrained \textit{request zone} to be completed by CNs distributed across an unbounded geographical~\textit{service zone}, thereby enhancing access to more computing power. We have derived analytical expressions to characterize the task completion probability as the main performance metric for this study. Moreover, we have examined performance optimization by managing both communication and propulsion energy consumption in practical hybrid fuel cell and battery-powered UAV scenarios. Extensive numerical results have validated the analytical results and highlighted the importance of balanced resource coordination to achieve optimal system performance. By effectively managing both communication parameters (e.g., UAV altitude and transmit power) and computing power parameters (i.e., computing capabilities, CN density, and CN distribution radius), we can significantly enhance the task completion probability. This work lays the foundation for future research on the integration of advanced computing and communication technologies into future-generation wireless networks, leading to the emerging computing power networks that have been considered indispensable for future AI-enabled applications.

\bibliography{deng}

@ARTICLE{Deng2024,
  author    = {Y. Deng and H. Zhang and X. Chen and Y. Fang},
  journal   = {IEEE Trans. Wireless Commun.}, 
  title     = {{UAV}-Assisted Multi-Access Edge Computing With Altitude-Dependent Computing Power}, 
  year      = {Aug. 2024},
  doi       = {10.1109/TWC.2024.3362375},
  note      = {23(8): 9404--9418}
}

@ARTICLE{Dong2024,
  author    = {W.-Y. Dong and S. Yang and P. Zhang and S. Chen},
  journal   = {IEEE J. Sel. Areas Commun.}, 
  title     = {Stochastic Geometry Based Modeling and Analysis of Uplink Cooperative Satellite-Aerial-Terrestrial Networks for Nomadic Communications With Weak Satellite Coverage}, 
  year      = {Dec. 2024},
  doi       = {10.1109/JSAC.2024.3459268},
  note      = {42(12): 3428--3444}
}

@ARTICLE{Fang2025,
  author    = {Z. Fang and S. Hu and J. Wang and Y. Deng and X. Chen and Y. Fang},
  journal   = {IEEE Trans. Netw.}, 
  title     = {Prioritized Information Bottleneck Theoretic Framework With Distributed Online Learning for Edge Video Analytics}, 
  year      = {Jan. 2025},
  pages     = {1--17},
  note      = {early access},
  doi       = {10.1109/TON.2025.3526148},
}

@ARTICLE{Hu2025,
  author    = {S. Hu and Z. Fang and Y. Deng and X. Chen and Y. Fang and S. Kwong},
  journal   = {IEEE Trans. Intell. Transp. Syst.}, 
  title     = {Toward Full-Scene Domain Generalization in Multi-Agent Collaborative Bird’s Eye View Segmentation for Connected and Autonomous Driving}, 
  year      = {Feb. 2025},
  note      = {26(2): 1783--1796},
  doi       = {10.1109/TITS.2024.3506284},
}

@article{Lalley20XX,
  author    = {S. P. Lalley},
  title     = {Poisson Processes},
  journal   = {Univ. Chicago, Dep. Statist.},
  url       = {https://galton.uchicago.edu/~lalley/Courses/312/PoissonProcesses.pdf},
}

@ARTICLE{al2014optimal,
  author={A. Al-Hourani and S. Kandeepan and S. Lardner},
  journal={IEEE Wireless Commun. Lett.}, 
  title={Optimal {LAP} Altitude for Maximum Coverage}, 
  year={Dec. 2014},
  note={3(6): 569--572}
}

@article{Møller_Schoenberg_2010,
  author    = {J. Møller and F. P. Schoenberg},
  title     = {Thinning spatial point processes into Poisson processes},
  journal   = {Adv. Appl. Probab.},
  year      = {Jul. 2010},
  note      = {42(2): 347--358},
  doi       = {10.1239/aap/1275055232},
}

@ARTICLE{Sun2024,
  author    = {Y. Sun and B. Lei and J. Liu and H. Huang and X. Zhang and J. Peng and W. Wang},
  journal   = {China Commun.}, 
  title     = {Computing power network: a survey}, 
  year      = {Sept. 2024},
  note      = {21(9): 109--145},
  doi       = {10.23919/JCC.ja.2021-0776},
}

@ARTICLE{Huang2018,
  author    = {S.-W. Ko and K. Han and K. Huang},
  journal   = {IEEE Trans. Wireless Commun.}, 
  title     = {Wireless Networks for Mobile Edge Computing: Spatial Modeling and Latency Analysis}, 
  year      = {Aug. 2018},
  note      = {17(8): 5225--5240}
}

@article{ma2025uav,
  title={{UAV}-Assisted Computing Power Network Task Allocation and {3D} Urban Trajectory Optimization},
  author={Ma, Bo and Pan, Yexin and Xu, Yong and Gao, Ziyi and Zhang, Zitian and Chen, Chao and Li, Chuanhuang},
  journal={IEEE Internet Things J.},
  year={Feb. 2025},
  note = {early access},
  publisher={IEEE}
}

@ARTICLE{tao2024multi,
  author={Tao, Ming and Li, Xueqiang and Feng, Jie and Lan, Dapeng and Du, Jun and Wu, Celimuge},
  journal={IEEE J. Sel. Areas Commun.}, 
  title={Multi-Agent Cooperation for Computing Power Scheduling in {UAV}s Empowered Aerial Computing Systems}, 
  year={Sept. 2024},
  note      = {42(12): 3521--3535}
}

@ARTICLE{deng2024uav,
  author    = {Deng, Yiqin and Zhang, Haixia and Chen, Xianhao and Fang, Yuguang},
  journal   = {IEEE Wirel. Commun.},
  title     = {{UAV}-Assisted {MEC} with an Expandable Computing Resource Pool: Rethinking the {UAV} Deployment},
  year      = {Oct. 2024},
  note      = {31(5): 110--116}
}

@ARTICLE{zhang2020energy,
  author    = {T. Zhang and G. Liu and H. Zhang and W. Kang and G. K. Karagiannidis and A. Nallanathan},
  journal   = {IEEE Trans. Commun.},
  title     = {Energy-Efficient Resource Allocation and Trajectory Design for {UAV} Relaying Systems},
  year      = {Oct. 2020},
  note      = {68(10): 6483--6498}
}

@ARTICLE{guo2025integrated,
  author    = {Xiaoyu Guo and Xiaowei Song and Dan Zeng and Zhen Dong and Xiang Yu and Lu Liu and Yuguang Fang},
  journal   = {IEEE/ASME Trans. Mechatron.},
  title     = {Integrated Energy-Efficient Planning and Management Framework for Autonomous Long-Endurance Flight of Hydrogen Fuel Cell/Battery Hybrid {UAV}s},
  year      = {Jan. 2025},
  note      = {early access}
}

@ARTICLE{xu20243d,
  author={Y. Xu and T. Zhang and Y. Liu and D. Yang and L. Xiao and M. Tao},
  journal={IEEE Trans. Veh. Technol.}, 
  title={{3D} Multi-{UAV} Computing Networks: Computation Capacity and Energy Consumption Tradeoff}, 
  year={Jul. 2024},
  note={73(7): 10627--10641}
}

@ARTICLE{qi2024minimizing,
  author={S. Qi and B. Lin and Y. Deng and X. Chen and Y. Fang},
  journal={IEEE Trans. Veh. Technol.}, 
  title={Minimizing Maximum Latency of Task Offloading for Multi-{UAV}-Assisted Maritime Search and Rescue}, 
  year={Sept. 2024},
  note={73(9): 13625--13638}
}

@ARTICLE{telikani2025unmanned,
  author={A. Telikani and A. Sarkar and B. Du and F. Santoso and J. Shen and J. Yan and J. Yong and E. Yap},
  journal={IEEE Commun. Surveys Tuts.}, 
  title={Unmanned Aerial Vehicle-Aided Intelligent Transportation Systems: Vision, Challenges, and Opportunities}, 
  year={Jan. 2025},
  note={early access}
}

@ARTICLE{liu2025multi,
  author={S. Liu and H. Yang and M. Zheng and L. Xiao},
  journal={IEEE Trans. Mobile Comput.}, 
  title={Multi-{UAV}-Assisted {MEC} in {IoV} with Combined Multi-Modal Semantic Communication under Jamming Attacks}, 
  year={Mar. 2025},
  note={early access}
}

@ARTICLE{xiao2025star,
  author={H. Xiao and X. Hu and W. Wang and Z. Su and K. Wong and K. Yang},
  journal={IEEE Trans. Commun.}, 
  title={{STAR-RIS} and {UAV} Combination in {MEC} Networks: Simultaneous Task Offloading and Communications}, 
  year={Jan. 2025},
  note={early access}
}

@ARTICLE{wang2024joint,
  author={C. Wang and D. Zhai and R. Zhang and L. Cai and L. Liu and M. Dong},
  journal={IEEE Trans. Veh. Technol.}, 
  title={Joint Association, Trajectory, Offloading, and Resource Optimization in Air and Ground Cooperative {MEC} Systems}, 
  year={Sept. 2024},
  note={73(9): 13076--13089}
}

@ARTICLE{nguyen2025integrated,
  author={M. D. Nguyen and W. Ajib and W. Zhu and G. Karabulut Kurt},
  journal={IEEE Trans. Wireless Commun.}, 
  title={Integrated User Association, Computation Offloading, Resource Allocation, and {UAV} Trajectory Control Against Jamming for {UAV}-Based Wireless Networks}, 
  year={Mar. 2025},
  note={early access}
}

@ARTICLE{nabi2025joint,
  author={A. Nabi and S. Moh},
  journal={IEEE Trans. Mobile Comput.}, 
  title={Joint Offloading Decision, User Association, and Resource Allocation in Hierarchical Aerial Computing: Collaboration of {UAV}s and {HAP}}, 
  year={Mar. 2025},
  note={early access}
}

@ARTICLE{lu2024resource,
  author={F. Lu and G. Liu and W. Lu and Y. Gao and J. Cao and N. Zhao and A. Nallanathan},
  journal={IEEE Trans. Commun.}, 
  title={Resource and Trajectory Optimization for {UAV}-Relay-Assisted Secure Maritime {MEC}}, 
  year={Mar. 2024},
  note={72(3): 1641--1652}
}

@ARTICLE{wu2025joint,
  author={T. Wu and M. Li and Y. Qu and H. Wang and Z. Wei and J. Cao},
  journal={IEEE Trans. Cognitive Commun. Netw.}, 
  title={Joint {UAV} Deployment and Edge Association for Energy-Efficient Federated Learning}, 
  year={Feb. 2025},
  note={early access}
}

@ARTICLE{zhu2024multi,
  author={X. Zhu and L. Zhai and N. Li and Y. Li and F. Yang},
  journal={IEEE Trans. Commun.}, 
  title={Multi-Objective Deployment Optimization of {UAV}s for Energy-Efficient Wireless Coverage}, 
  year={Jun. 2024},
  note={72(6): 3587--3601}
}

@ARTICLE{gong2024energy,
  author={H. Gong and B. Huang and B. Jia},
  journal={IEEE Trans. Mobile Comput.}, 
  title={Energy-Efficient 3-D {UAV} Ground Node Accessing Using the Minimum Number of {UAV}s}, 
  year={Dec. 2024},
  note={23(12): 12046--12060}
}

@ARTICLE{lin2024a,
  author={X. Lin and S. Bi and G. Su and Y. Zhang},
  journal={IEEE Internet Things J.}, 
  title={A Lyapunov-Based Approach to Joint Optimization of Resource Allocation and 3-D Trajectory for Solar-Powered {UAV} {MEC} Systems}, 
  year={Jun. 2024},
  note={11(11): 20797--20815}
}

@ARTICLE{tang2021computing,
  author={X. Tang and C. Cao and Y. Wang and S. Zhang and Y. Liu and M. Li and T. He},
  journal={China Commun.}, 
  title={Computing Power Network: The Architecture of Convergence of Computing and Networking Towards 6G Requirement}, 
  year={Feb. 2021},
  note={18(2): 175--185}
}

@ARTICLE{basu2019a,
  author={S. Basu and S. Roy and S. DasBit},
  journal={IEEE Trans. Eng. Manag.}, 
  title={A Post-Disaster Demand Forecasting System Using Principal Component Regression Analysis and Case-Based Reasoning Over Smartphone-Based DTN}, 
  year={May 2019},
  note={66(2): 224--239}
}

@ARTICLE{pan2024resource,
  author={W. Pan and N. Lv and B. Hou and Z. Ren},
  journal={IEEE Trans. Netw. Sci. Eng.}, 
  title={Resource Allocation and Outage Probability Optimization Method for Multi-Hop {UAV} Relay Network for Servicing Heterogeneous Users}, 
  year={Jan. 2024},
  note={11(3): 2769--2781}
}

@INPROCEEDINGS{deng2025uav_c,
  author = {Deng, Yiqin and Fang, Zhengru and Hu, Senkang and Ma, Yanan and Zhang, Haixia and Fang, Yuguang},
  booktitle = {IEEE GLOBECOM},
  title = {{UAV}-enabled Computing Power Networks: Task Completion Probability Analysis},
  year = {Taipei, Taiwan, Dec. 8-12, 2025}
}

@ARTICLE{deng2022actions,
  author={Y. Deng and X. Chen and G. Zhu and Y. Fang and Z. Chen and X. Deng},
  journal={IEEE Wireless Commun.}, 
  title={Actions at the Edge: Jointly Optimizing the Resources in Multi-Access Edge Computing}, 
  year={Apr. 2022},
  note={29(2): 192--198}
}

@ARTICLE{pervez2024energy,
  author={F. Pervez and A. Sultana and C. Yang and L. Zhao},
  journal={IEEE Trans. Wireless Commun.}, 
  title={Energy and Latency Efficient Joint Communication and Computation Optimization in a Multi-{UAV}-Assisted {MEC} Network}, 
  year={Mar. 2024},
  note={23(3): 1728--1741}
}

@ARTICLE{wang2023recent,
  author={X. Wang and Y. Jin and S. Schmitt and M. Olhofer},
  journal={ACM Computing Surveys}, 
  title={Recent Advances in {Bayesian} Optimization}, 
  year={Jul. 2023},
  note={55(13s): 1--36}
}

@ARTICLE{dai2024uav,
  author={X. Dai and Z. Xiao and H. Jiang and J. C. S. Lui},
  journal={IEEE Trans. Mobile Comput.}, 
  title={{UAV}-Assisted Task Offloading in Vehicular Edge Computing Networks}, 
  year={Apr. 2024},
  note={23(4): 2520--2534}
}

@ARTICLE{hao2024joint,
  author={H. Hao and C. Xu and W. Zhang and S. Yang and G.-M. Muntean},
  journal={IEEE Trans. Mobile Comput.}, 
  title={Joint Task Offloading, Resource Allocation, and Trajectory Design for Multi-{UAV} Cooperative Edge Computing With Task Priority}, 
  year={Sept. 2024},
  note={23(9): 8649--8663}
}

@article{ma2025raise,
  title={RAISE: Optimizing RIS Placement to Maximize Task Throughput in Multi-Server Vehicular Edge Computing},
  author={Ma, Yanan and Fang, Zhengru and Yuan, Longzhi and Deng, Yiqin and Chen, Xianhao and Fang, Yuguang},
  journal={arXiv preprint arXiv:2503.17708},
  year={2025}
}

@ARTICLE{Xu2018UAV,
  author={Xu, Jie and Zeng, Yong and Zhang, Rui},
  journal={IEEE Transactions on Wireless Communications}, 
  title={UAV-Enabled Wireless Power Transfer: Trajectory Design and Energy Optimization}, 
  year={2018},
  volume={17},
  number={8},
  pages={5092-5106}
}

@ARTICLE{Zeng2017Energy,
  author={Zeng, Yong and Zhang, Rui},
  journal={IEEE Transactions on Wireless Communications}, 
  title={Energy-Efficient UAV Communication With Trajectory Optimization}, 
  year={2017},
  volume={16},
  number={6},
  pages={3747-3760},
}

@ARTICLE{Zeng2019Energy,
author={Zeng, Yong and Xu, Jie and Zhang, Rui},
journal={IEEE Transactions on Wireless Communications},
title={Energy Minimization for Wireless Communication With Rotary-Wing UAV},
year={2019},
volume={18},
number={4},
pages={2329-2345}
}

@ARTICLE{Zhan2018Energy,
author={Zhan, Cheng and Zeng, Yong and Zhang, Rui},
journal={IEEE Wireless Communications Letters},
title={Energy-Efficient Data Collection in UAV Enabled Wireless Sensor Network},
year={2018},
volume={7},
number={3},
pages={328-331}
}

@ARTICLE{Xiao2024Space,
author={Xiao, Yue and Ye, Ziqiang and Wu, Mingming and Li, Haoyun and Xiao, Ming and Alouini, Mohamed-Slim and Al-Hourani, Akram and Cioni, Stefano},
journal={IEEE Journal on Selected Areas in Communications},
title={Space-Air-Ground Integrated Wireless Networks for 6G: Basics, Key Technologies, and Future Trends},
year={Dec. 2024},
volume={42},
number={12},
pages={3327-3354}
}

@article{zhao2025dynamic,
  title={Dynamic-Memory Event-Triggered Fixed-Time Distributed Power Management for {UAV} Power System under Intermittent Communication Failures},
  author={Zhao, Sailiu and Ni, Junkang and Lei, Tao and Du, Yuhau and Deng, Shuhao},
  journal={IEEE Transactions on Aerospace and Electronic Systems},
  year={July 2025},
  note={early access}
}

@misc{lyu2025empowering,
      title={Empowering Intelligent Low-altitude Economy with Large AI Model Deployment}, 
      author={Zhonghao Lyu and Yulan Gao and Junting Chen and Hongyang Du and Jie Xu and Kaibin Huang and Dong In Kim},
      year={2025},
      eprint={2505.22343},
      archivePrefix={arXiv},
      primaryClass={eess.SP},
      url={https://arxiv.org/abs/2505.22343}, 
}

@ARTICLE{Li2025Large,
  author={Li, Haoyun and Xiao, Ming and Wang, Kezhi and Kim, Dong In and Debbah, Merouane},
  journal={IEEE Wireless Communications Letters}, 
  title={Large Language Model Based Multi-Objective Optimization for Integrated Sensing and Communications in UAV Networks}, 
  year={April 2025},
  volume={14},
  number={4},
  pages={979-983}
}

@ARTICLE{Li2023Channel,
  author={Li, Haoyun and Li, Peiming and Cheng, Gaoyuan and Xu, Jie and Chen, Junting and Zeng, Yong},
  journal={Journal of Communications and Information Networks}, 
  title={Channel Knowledge Map (CKM)-Assisted Multi-UAV Wireless Network: CKM Construction and UAV Placement}, 
  year={Sept 2023},
  volume={8},
  number={3},
  pages={256-270}
}

\begin{IEEEbiography}[{\includegraphics[width=1in,height=1.25in,clip,keepaspectratio]{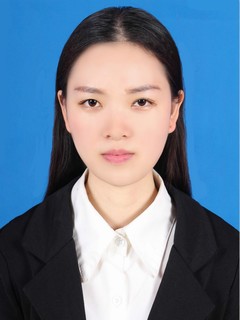}}]{Yiqin Deng}(Member, IEEE) received the M.S. degree in software engineering and the Ph.D. degree in computer science and technology from Central South University, Changsha, China, in 2017 and 2022, respectively. She is currently a Research Assistant Professor with the School of Data Science, Lingnan University, Hong Kong. From 2024 to 2026, she was a Postdoctoral Research Fellow in the Department of Computer Science, City University of Hong Kong. Prior to that, she was a Postdoctoral Research Fellow with the School of Control Science and Engineering, Shandong University, Jinan, China, from 2022 to 2024. She also served as a Visiting Researcher at the University of Florida, Gainesville, FL, USA, from 2019 to 2021. Her research interests include edge computing/AI, wireless communication and networking, computing power networks, and the low-altitude economy.
\end{IEEEbiography}

\begin{IEEEbiography}[{\includegraphics[width=1in,height=1.25in,clip,keepaspectratio]{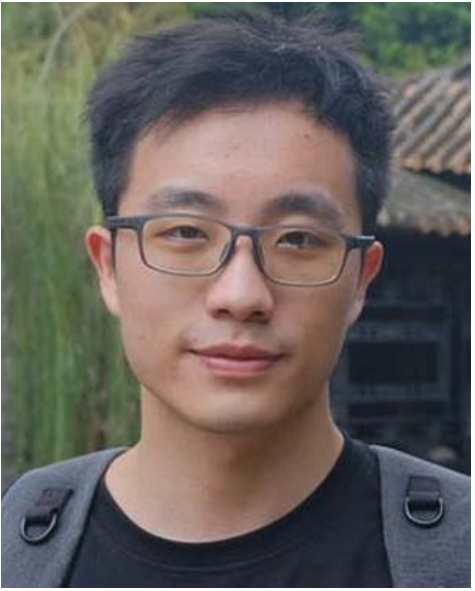}}]{Zhengru Fang}
(S'20) received his B.S. degree (Hons.) in electronics and information engineering from the Huazhong University of Science and Technology (HUST), Wuhan, China, in 2019 and received his M.S. degree (Hons.) from Tsinghua University, Beijing, China, in 2022. Currently, he is pursuing his PhD degree in the Department of Computer Science at City University of Hong Kong, where he is also a recipient of the Hong Kong PhD Fellowship Scheme (HKPFS). His research interests include collaborative perception, V2X, age of information, and mobile edge computing. He received the Outstanding Thesis Award from Tsinghua University in 2022, and the Excellent Master Thesis Award from the Chinese Institute of Electronics in 2023. His research work has been published in IEEE/CVF CVPR, IEEE ToN, IEEE JSAC, IEEE TMC, IEEE ICRA, and ACM MM, etc.
\end{IEEEbiography}

\begin{IEEEbiography}[{\includegraphics[width=1in,height=1.25in,clip,keepaspectratio]{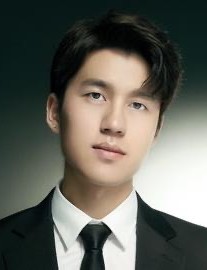}}]{Senkang Hu} 
         is currently a Ph.D. candidate at the Department of Computer Science, City University of Hong Kong and the Hong Kong JC STEM Lab of Smart City. Previously, he received his BEng degree from the School of Information and Electronics, Beijing Institute of Technology in 2022. He focuses on LLM-empowered multi-agent systems, LLM post-training, and autonomous driving. His works have been published in top-tier journals such as TMC, TITS, ToN, and top-tier conferences such as AAAI, ICRA, etc. He  also serves as a reviewer or technical program committee member for ICML, ICLR, NeurIPS, ICRA, TMC, ToN, TITS, etc.
\end{IEEEbiography}

\begin{IEEEbiography}[{\includegraphics[width=1in,height=1.25in,clip,keepaspectratio]{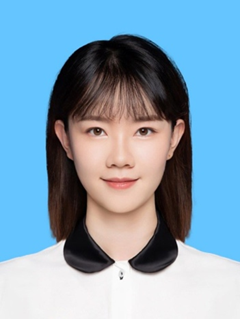}}]{Yanan Ma}(Graduate Student Member, IEEE) received the B.Eng. degree (Hons.) in Electronic Information Engineering (English Intensive) and the M.Eng. degree (Hons.) in Information and Communication Engineering from the Dalian University of Technology, Dalian, China, in 2020 and 2023, respectively. She is currently pursuing the Ph.D. degree in the Department of Computer Science at the City University of Hong Kong. She received the IEEE GLOBECOM Best Paper Award in 2025. Her research interests are focused on edge intelligence, wireless communication and networking, and machine learning.\end{IEEEbiography}

\begin{IEEEbiography}[{\includegraphics[width=1in,height=1.25in,clip,keepaspectratio]{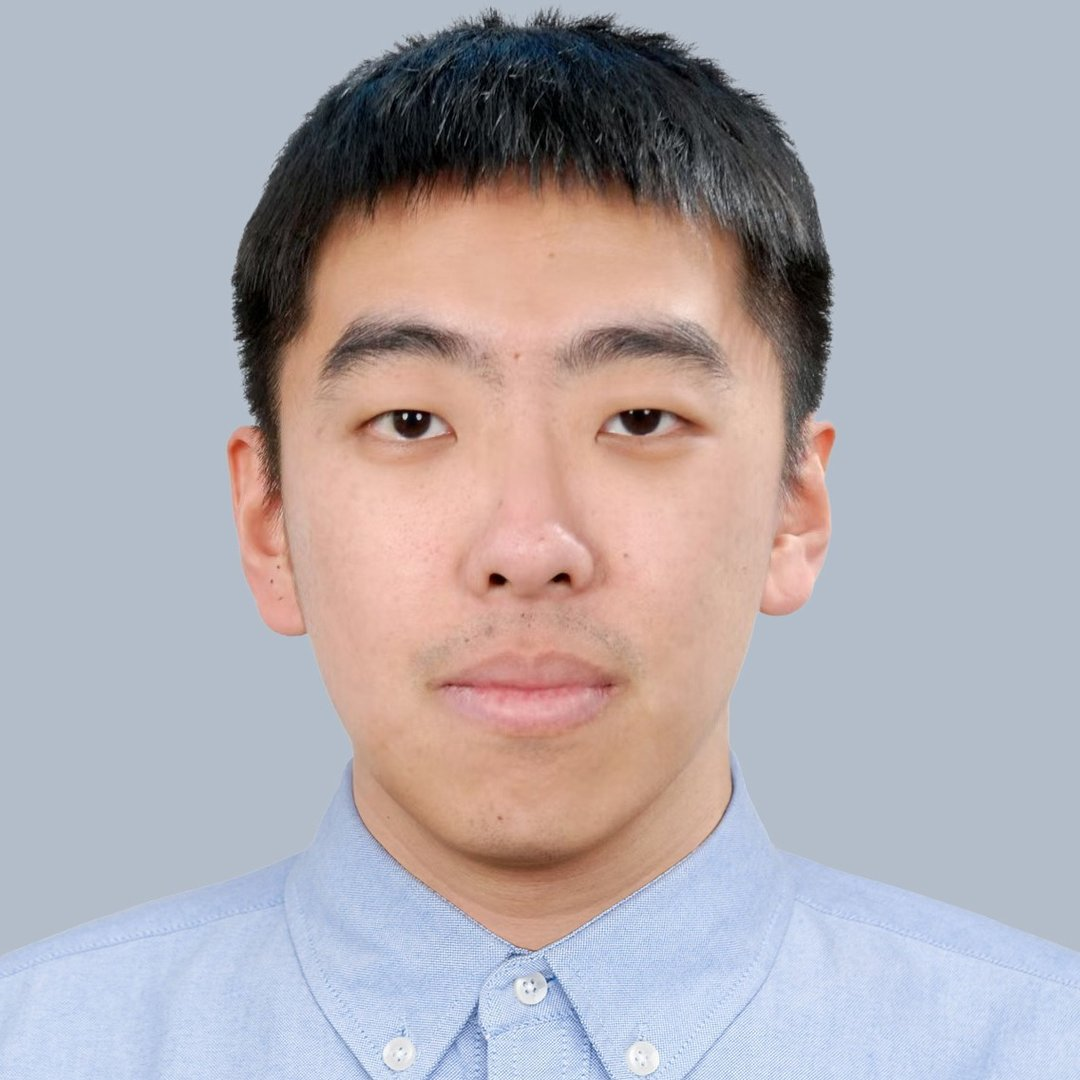}}]{Xiaoyu Guo} received his Bachelors degree from Beihang University in 2018, Masters Degree (with Distinction) from the University of Cambridge in 2019, and PhD from the University of Manchester in 2023. He is currently an assistant professor with the Department of Mechanical Engineering, City University of Hong Kong.

His research interests include hydrogen-powered robotic systems, and bio-inspired control and perception. He has published over 40 papers in international journals including Nature Communications, IEEE Transactions on Automatic Control, Automatica, and IEEE/ASME Transactions on Mechatronics. He has also been also granted more than 20 invention patents. He serves as an assosciate editor for IEEE Transactions on Industrial Informatics, and Unmanned Systems.
\end{IEEEbiography}

\begin{IEEEbiography}[{\includegraphics[width=1in,height=1.25in,clip,keepaspectratio]{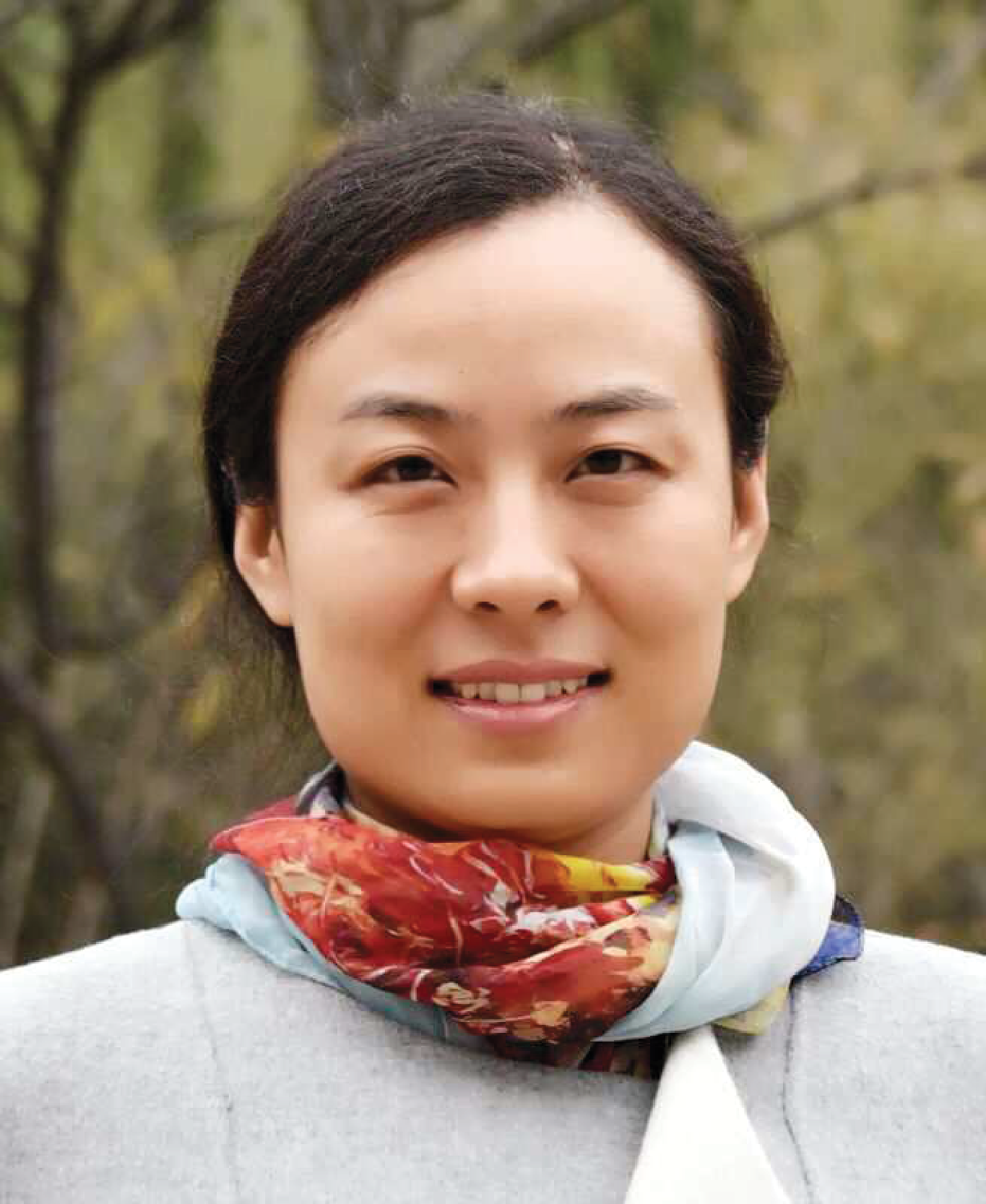}}]{Haixia Zhang}(M'08-SM'11) received the B.E. degree from the Department of Communication and Information Engineering, Guilin University of Electronic Technology, Guilin, China, in 2001, and the M.Eng. and Ph.D. degrees in communication and information systems from the School of Information Science and Engineering, Shandong University, Jinan, China, in 2004 and 2008, respectively. 

From 2006 to 2008, she was with the Institute for Circuit and Signal Processing, Munich University of Technology, Munich, Germany, as an Academic Assistant. From 2016 to 2017, she was a Visiting Professor with the University of Florida, Gainesville, FL, USA. She is currently a Full Professor with Shandong University, Jinan, China. Dr. Zhang is actively participating in many professional services. She is/was an editor of the IEEE Transactions on Wireless Communications, IEEE Internet of Things Journal, IEEE Wireless Communication Letters, and China Communications and serves/served as  Symposium Chairs, TPC Members, Session Chairs, and Keynote Speakers of many conferences. Her research interests include wireless communication and networks, industrial Internet of Things, wireless resource management, and mobile edge computing. 
\end{IEEEbiography}

\begin{IEEEbiography}[{\includegraphics[width=1.2in,height=1.3in,clip,keepaspectratio]{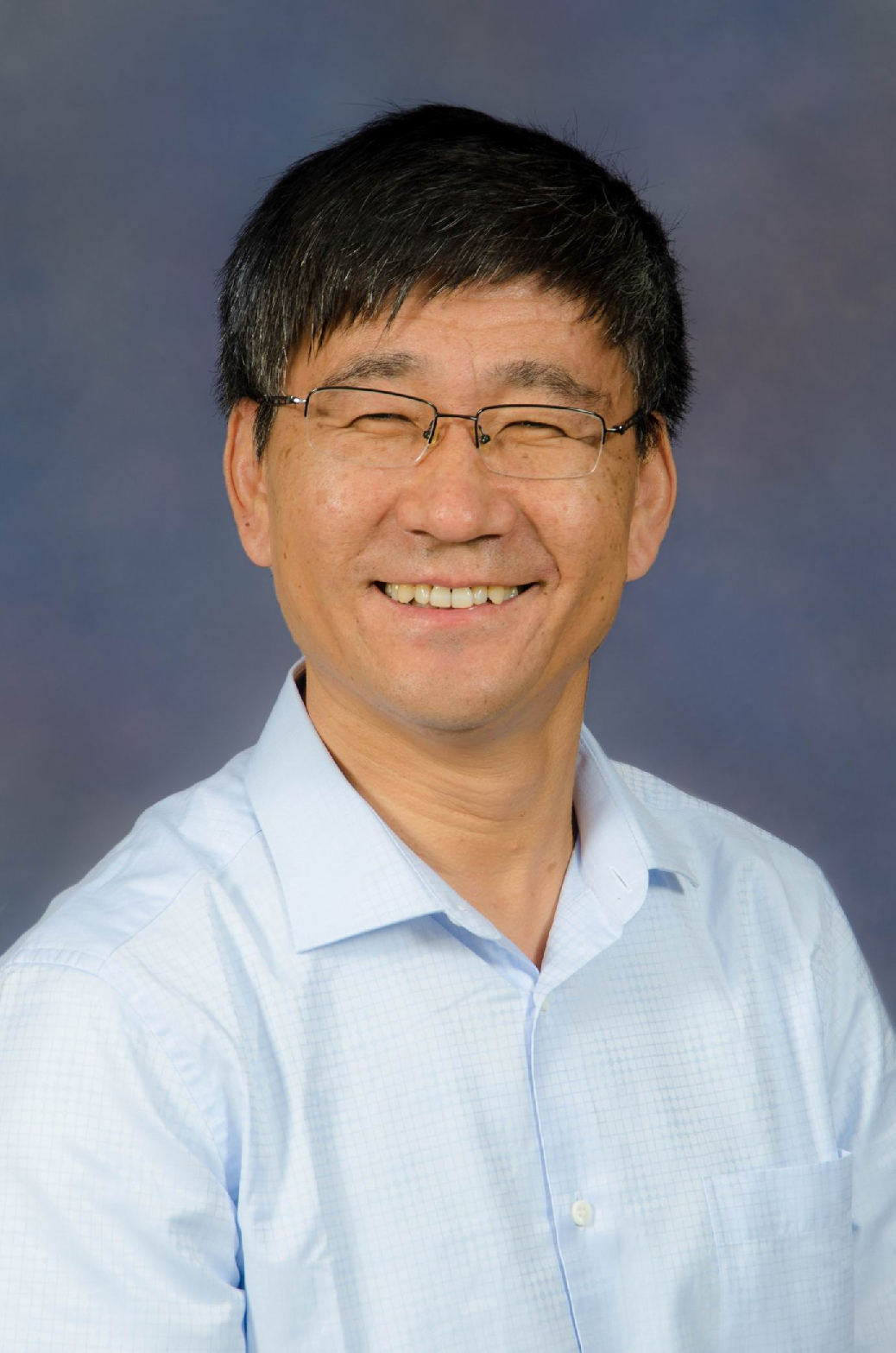}}]{Yuguang Fang} (S’92, M’97, SM’99, F’08) received an MS degree from Qufu Normal University, China, a PhD degree from Case Western Reserve University, USA, and another PhD degree from Boston University, USA, in 1987, 1994, and 1997, respectively. He joined the Department of Electrical and Computer Engineering at University of Florida in 2000 as an assistant professor, then was promoted to associate professor, full professor, and distinguished professor, in 2003, 2005, and 2019, respectively. Since 2022, he has been a Global STEM Scholar and Chair Professor with Department of Computer Science, City University of Hong Kong. He is the Founding Director of Hong Kong JC STEM Lab of Smart City funded by The Hong Kong Jockey Club Charities Trust. 

He received many awards including US NSF CAREER Award, US ONR Young Investigator Award, the 2018 IEEE Vehicular Technology Outstanding Service Award, and several IEEE Communications Society awards (AHSN Technical Achievement Award, CISTC Technical Recognition Award, and WTC Recognition Award). He was the Editor-in-Chief of IEEE Transactions on Vehicular Technology and IEEE Wireless Communications. He is a fellow of ACM and AAAS.  
\end{IEEEbiography}

\end{document}